\documentclass{svjour3}

\usepackage[utf8]{inputenc}
\usepackage[dvips]{graphicx}
\usepackage{amscd}
\usepackage{amsmath}
\usepackage{amssymb}
\usepackage{amstext}
\usepackage{url}
\usepackage{epsfig}
\usepackage{wasysym}
\usepackage{fancyvrb}
\usepackage{alltt}
\usepackage{hhline}
\usepackage{psfrag}
\usepackage{array}
\usepackage{setspace}
\usepackage{float}
\usepackage{multirow}

\usepackage{ulem}

\usepackage{pst-tree}

\usepackage[sort&compress]{natbib} 
\bibpunct{(}{)}{;}{a}{}{,}

\def\K{\mathcal{K}}
\def\N{\mathbb{N}}

\def\Deg{^{\circ}}

\def\r{\boldsymbol{r}}
\def\v{\boldsymbol{v}}
\def\exc{\boldsymbol{e}}
\def\n{\boldsymbol{n}}
\def\j{\boldsymbol{j}}
\def\np{\boldsymbol{n}_p}
\def\ns{\boldsymbol{n}_{_{3b}}}
\def\es{e_{_{3b}}}
\def\epsp{\varepsilon_{_{\!J2}}}
\def\epsJ3{\varepsilon_{_{\!J3}}}
\def\epss{\varepsilon_{_{3b}}}

\newcommand{\dpar}[2]{\frac{\partial #1}{\partial #2}}
\newcommand{\jac}[3]{\frac{\partial^2 #1}{\partial #2\;\partial #3}}
\newcommand{\jacc}[2]{\frac{\partial^2 #1}{{\partial #2}^2}}
\newcommand{\ateq}[1]{\text{Eq.(\ref{#1})}}

\graphicspath{{/EPSF/}{Figures/}{./}}

\begin{document}


  \title{Frozen Orbits at high eccentricity and inclination: Application to Mercury orbiter.}
  \author{N. Delsate$^1$ \and 
  P. Robutel$^2$ \and 
  A. Lema\^{i}tre$^1$ \and
  T. Carletti$^1$}
\institute{$^1$ University of Namur, Department of Mathematics 
  Rempart de la Vierge 8, B-5000 NAMUR,BELGIUM \\
  \email{nicolas.delsate@math.fundp.ac.be} \\
  \and 
  $^2$ IMCCE, CNRS UMR8028, Observatoire de Paris/USTL/UPMC, 
  77 Av. Denfert-Rochereau, F-75014 Paris, France.
}
\date{Received: date / Accepted: date}

\maketitle
  \begin{abstract}
    We hereby study the stability of a massless probe orbiting 
    around an oblate central body (planet or planetary satellite) 
    perturbed by a third body, assumed to lie in the equatorial 
    plane (Sun or Jupiter for example) using an Hamiltonian 
    formalism. 

    We are able to determine, in the parameters space, the location of 
    the frozen orbits, namely orbits whose orbital elements remain 
    constant on average, to characterize their stability/unstability and 
    to compute the periods of the equilibria. 

    The proposed theory is general enough, to be applied to a wide 
    range of probes around planet or natural planetary satellites. 

    The BepiColombo mission is used to motivate our analysis and to 
    provide specific numerical data to check our 
    analytical results. 

    Finally, we also bring to the light that the 
    coefficient $J_2$ is able to protect against the increasing 
    of the eccentricity due to the Kozai-Lidov effect.


  \end{abstract}
  


\section{Introduction}

BepiColombo (MPO and MMO orbiters) is a joint European and 
Japanese space agencies space mission aimed at studying the 
planet Mercury. The MPO (Mercury Planetary Orbiter) 
will be brought into a polar elliptical orbit around 
Mercury with an inclination of $88–90\Deg$, an eccentricity of 
$0.1632$ and a semi-major axis of $3\,394$ km. The MMO 
(Mercury Magnetospheric Orbiter) will also be brought into 
a polar elliptical orbit with 
an eccentricity of $0.6679$ and a semi-major axis of $8\,552$ km. 

Actually polar orbits are very interesting for scientific 
missions to planetary satellites (with near polar low-altitude) 
or to planet (with high-eccentric high-altitude). 
The orbital dynamics of such space probes is governed by the oblateness
($J_2$ effect) of the central body around which the space probe 
is orbiting and the gravity field 
from the third body. A well-known effect of the third-body 
perturbation is the change in the stability of circular orbits 
related to orbit inclination. This effect is a natural consequence 
of the Kozai-Lidov resonance (\citealt{Kozai1962,Lidov1963}). 
The final fate of such a satellite is the collision with the central 
body. Therefore the control of the orbital eccentricity leads 
to the control of the satellite lifetime.

\cite{Scheeres2001} studied near-circular 
orbits in a model that included both the third body's gravity 
and $J_2$. In addition 
\cite{SanJuan2006} studied orbit dynamics about oblate planetary 
satellites using a rigorous averaging method. \cite{Paskowitz2006} 
added the effect of the coefficient $J_3$. These authors mainly focused 
their attention to an orbiter around planetary satellites especially 
for Europa orbiter. So they did not take into account the 
eccentricity of the third body and they detailed the 
near-circular orbits.

Our purpose is to build a simplified Hamiltonian model, 
as simple as possible, which will reproduce the motion of probes 
orbiting an oblate central body also taking into account 
the third body effect. Especially we are looking for the conditions 
that give rise to frozen orbits. 
Frozen orbits are orbits that have orbital elements constant on average. 
These particular orbits are able to keep constant the eccentricity. 
Therefore in a neighbourhood of these orbits there is a stability area 
where even a limited control could be used to avoid the crash 
onto the central body.

Beside the oblateness of the central body and the gravity effect 
of the third body, our averaged model takes into account also the 
eccentricity of the orbit of the third body. 
Moreover let us observe that our results are given in closed form 
with respect to eccentricity and inclination of the probe, namely we do 
not perform any power series expansion; therefore, our theory 
applies for arbitrary eccentricities and inclinations of the 
space probe, and is not limited to almost-circular orbits. 
We can thus conclude that the theory is general 
enough to be applied to a wide range of probes around a planet or 
around a natural planetary satellite and, can be formulated and 
presented in a general way that allows extension of the 
results to other cases.

The Mercury orbiter mission (BepiColombo) is used to motivate 
our analysis and to provide specific numerical data to check 
our analytical results.

We are able to provide the location of frozen orbits and study 
their stability as a function of the involved parameters, 
using implicit equations and graphics. 
Finally we give the analytical expressions of the 
periods at the stable equilibria. 

The analytical results are verified and confirmed using dedicated 
numerical simulations of the whole model. 

To conclude, we discuss the effect of the protection of $J_2$ 
on the increase of the eccentricity due to Kozai-Lidov effect and the 
apparition of an asymmetry caused by the addition of the coefficient $J_3$.

\section{Motivation: numerical exploration}

For the purpose of our study, we consider the modeling of a space 
probe subjected to the influence of Mercury's gravity field 
(in the following sections Mercury will be denoted by ``central body'') and 
the gravitational perturbations of the Sun (noted ``third body'') 
as well as to the direct solar radiation pressure without 
shadowing effect. As a consequence the differential system of 
equations is given by 
\begin{equation}\label{diffEq}
  \ddot{\boldsymbol{r}} = \ddot{\boldsymbol{r}}_{\text{pot}} 
  + \ddot{\boldsymbol{r}}_{\odot} +  \ddot{\boldsymbol{r}}_{\text{rp}}\, ,
\end{equation}
where $\ddot{\boldsymbol{r}}_{\text{pot}}$ is the acceleration induced by 
Mercury's gravity field, $\ddot{\boldsymbol{r}}_{\odot}$ is the 
acceleration resulting from the gravity interaction with the 
Sun and $\ddot{\boldsymbol{r}}_{rp}$ is the acceleration due to the 
direct solar radiation pressure. 

It is worth noting that we modelise the gravity potential of central body  
only using the $J_2$, $C_{22}$ and $J_3$ coefficients. 
In our implementation, we choose the high accurate Solar System 
ephemeris given by the Jet Propulsion Laboratory (\texttt{JPL}) 
to provide the positions of the Sun \citep{standish98}. 
We adopt the variable step size Bulirsch-Stoer algorithm (see e.g. 
\citealt{bulirsh-stoer}) to numerically integrate the 
differential equation~(\ref{diffEq}). Let us note that, for the purpose 
of validation, we also use a second numerical integrator \texttt{DOP853}
(an explicit Runge-Kutta method of order 8(5,3) with stepsize control 
due to Dormand \& Prince \citep{Hairer1993}).

\begin{figure}[htbp] 
  \begin{center}
    \includegraphics[draft=false,width=\textwidth]{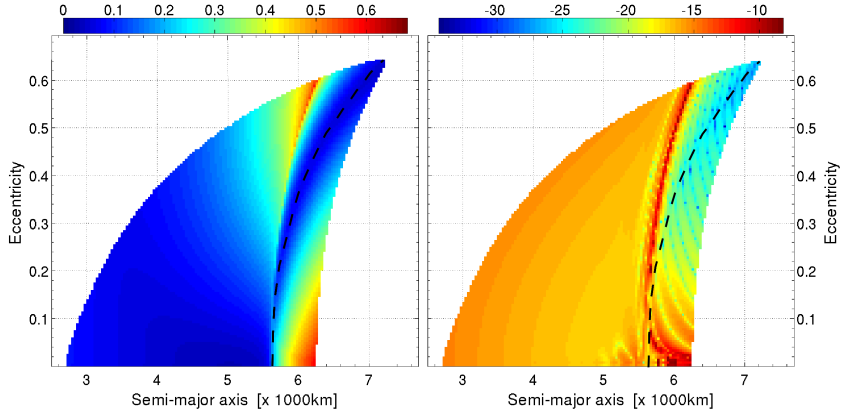}
    \caption{\label{excMap}The eccentricity computed as
      a function of the initial eccentricity $e_0$ and the initial
      semi-major axis $a_0$. The equations of motion
      include the central body attraction, the harmonics $J_2,C_{22},J_3$, 
      the solar interaction as well as the perturbing effects of the 
      solar radiation pressure ($A/m=0.01$m$^2$/kg). 
      The eccentricity step is $0.005$ and the
      semi-major axis step is $35$~km. The initial conditions are 
      $i_0 = 90\Deg,\Omega_0 = 67.7\Deg, \omega_0 =-2\Deg$ and $M_0=36.4\Deg$.
      The integration time is 200~years from epoch fixed at 
      14~September~2019. The patterns have been obtained by 
      plotting the amplitude of variation of the eccentricity and 
      $\log(\delta\delta\nu)$ respectively in left and right panel.}
  \end{center}
\end{figure}

In Figure~\ref{excMap} we report the results of a numerical 
integration of the system of equations (\ref{diffEq}) 
for a set of 19\,600 orbits, 
propagated over a 200~years time span with a entry-level step 
size of 300~seconds. 
We consider a set of initial conditions defined by an 
eccentricity grid of $0.005$ and a semi-major axis 
grid of 35~km, spanning the $[2600,7600]$~km range. 
The other fixed initial conditions are $i_0=90\Deg$ for the inclination, 
$\Omega_0=67.7\Deg$ $\omega_0=-2\Deg$ for the longitude of the 
ascending node and the argument of periherm, respectively; 
$M_0=36.4\Deg$ the mean anomaly at epoch fixed at 14 
September 2019. The area-to-mass ratio $A/m=0.01$m$^2$/kg. 
These values have been fixed by the initial 
conditions of BepiColombo mission found in \citet{Garcia2007}. 

We show the amplitude of the eccentricity (that is the difference between 
the maximum and minimum eccentricity reached during the integration) of 
each orbit in the left panel of Figure~\ref{excMap}. For each orbit, 
we also calculate using the Numerical Analysis of Fundamental 
Frequencies, for short \texttt{NAFF} \citep{Laskar1988,Laskar2005}, 
the fundamental frequency (noted $\nu$) of the evolution 
of the eccentricity vector ($e\cos\omega,e\sin\omega$). We plot 
the logarithm of the second derivative (noted $\log(\delta\delta\nu)$) 
of this frequency in the right panel of Figure~\ref{excMap}, 
namely an indicator of the diffusion in the frequency space, 
hence the regularity of the orbit. 
For more details concerning this use of frequency analysis, 
see \cite{lemaitre2009} where the frequency analysis has been 
used to study resonances in Geostationary Earth Orbits. 

First, let us observe that the white zone in Figure~\ref{excMap} 
corresponds to orbits that crash onto central body's surface. 
Second we distinguish a curve where the variation of the 
eccentricity amplitude is null (dashed black line). On the second 
derivative plot (right panel) we also distinguish on the left of 
the dashed black line (null-variation of eccentricity) a 
larger value of the log of the derivative that could correspond 
to a separatrix. 

These structures will be analyzed and explained using a 
simplified model, that takes into account the central body attraction 
with the $J_2$ harmonic coefficient and the third body gravitational 
effect. We observed that the solar radiation pressure 
does not play any role in these structures, hence this effect 
will be absent in the simplified model.

\section{The Hamiltonian Formalism}

The aim of this section is to introduce the Hamiltonian (\ref{HamTrem}) 
already found in \citet{Tremaine2009}. Kepler's Hamiltonian 
describing the motion of a test particule orbiting an 
isolated point mass M is 
\begin{equation*}
  \mathcal{H}_K = \frac{1}{2}v^2-\frac{GM}{r}=\frac{GM}{a}
\end{equation*}
where $G$ is the gravitational constant, $\r$ is the planetocentric 
position of the particule, $\v=\dot{\r}$, $r=|\r|$ and 
$a$ is the semi-major axis of the particule. 

\begin{figure}[!htb] 
  \begin{center}
    \includegraphics[draft=false,width=0.47\textwidth]{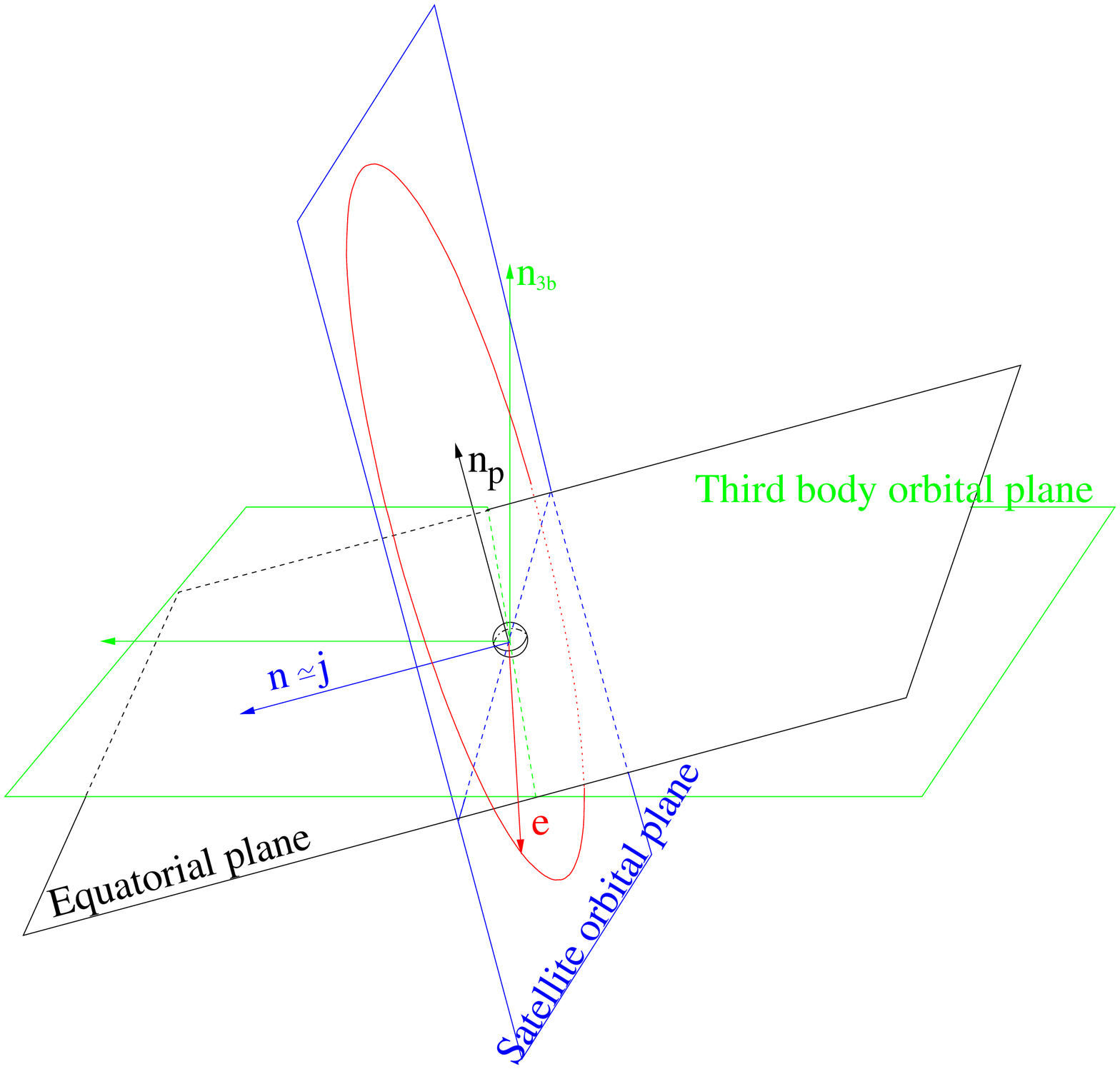}
    \includegraphics[draft=false,width=0.47\textwidth]{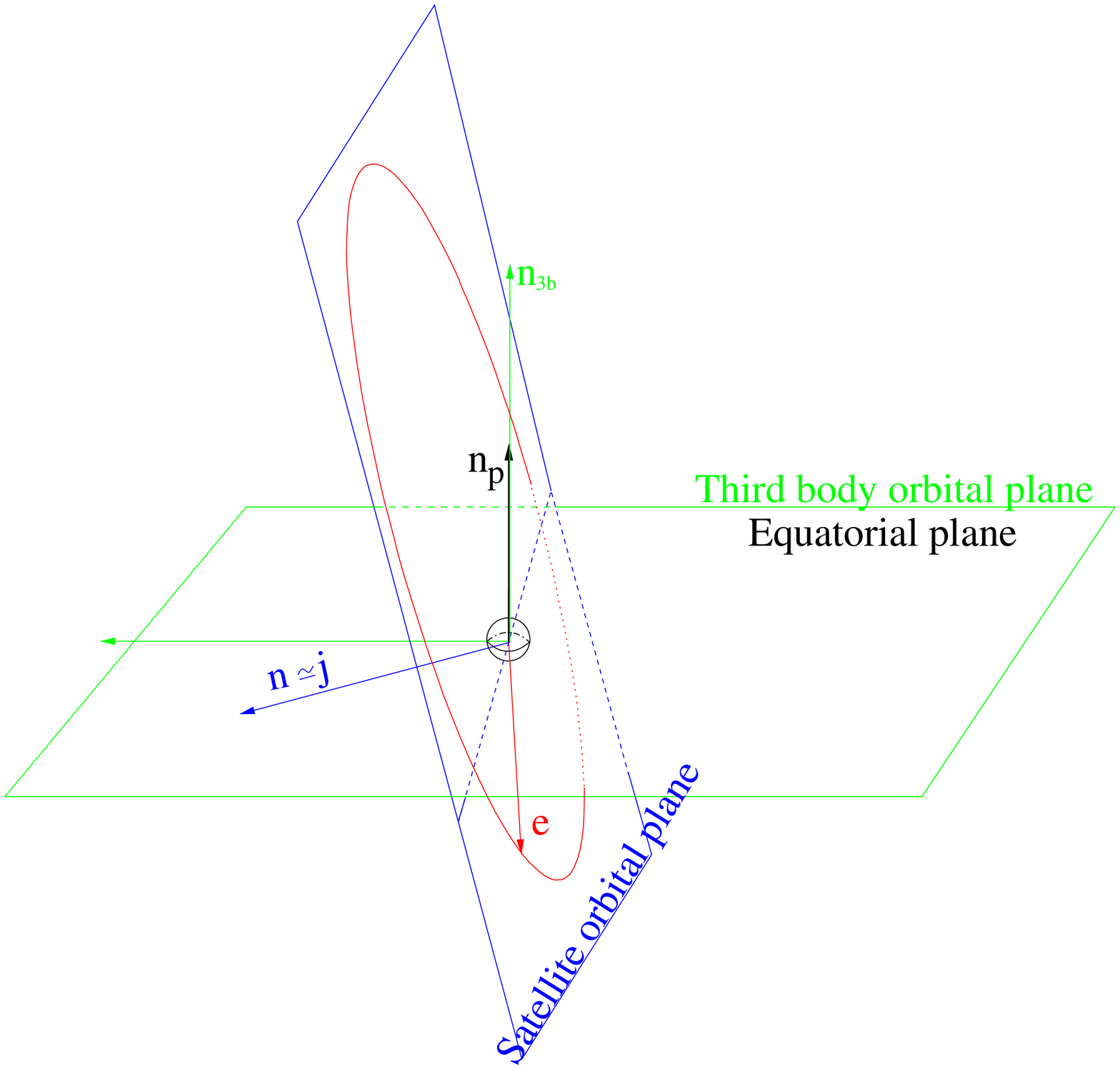}
    \caption{\label{laplaceSurf}Reference planes, for the general theory 
      on the left and for our particular case on the right.}
  \end{center}
\end{figure}

One can introduce the quadrupole potential arising from an oblate 
planet (``central body'') that is 
\begin{equation*}
  \Phi_{_{\!J2}}(\r) = \frac{GM J_2 R_p^2}{2r^5}\Big[3(\r\cdot\np)^2-r^2\Big]
\end{equation*}
where $\np$ is the unit vector oriented to central body's spin 
axis (see Figure~\ref{laplaceSurf}). 
$M$, $R_p$ and $J_2$ are respectively the mass, the radius and the 
oblateness coefficient of the central body (planet or natural satellite).

We assume that $r\ll a_{{_{3b}}}$ (where the subscript $_{3b}$ is related 
to the third body) and we average over the third body orbital period. 
So, we obtain the quadrupole in terms of the third body gravitational effect 
\begin{equation*}
  \Phi_{_{3b}}(\r)=\frac{GM_{{_{3b}}}}{4a_{{_{3b}}}^3(1-\es^2)^{3/2}}\Big[ 3(\r\cdot\ns)^2-r^2\Big]
\end{equation*}
where $\ns$ is the normal to the central body orbit. $M_{{_{3b}}}$, 
$a_{{_{3b}}}$ and $\es$ are respectively the mass, the semi-major axis and the 
eccentricity 
of the third body. This quadrupole term takes into account 
the eccentricity ($\es$) of the third body (e.g. Sun or Jupiter) 
around the central body (planet or natural satellite). Let us stress 
the fact that \cite{Scheeres2001,Paskowitz2004,SanJuan2006} 
do not include this eccentricity factor in their formulation, 
while for a Sun-Mercury-orbiter application, this 
will be an important contribution.

We then average over the Keplerian orbit of the test particule 
described by the following elements: a semi-major axis $a$, 
an eccentricity $e$, and an orientation specified by the unit 
vectors $\n$ along the angular momentum vector, $\boldsymbol{u}$ 
toward the pericenter and $\v=\n\times\boldsymbol{u}$. 
We have \citep{BrouwerClemence}
\begin{equation*}
  \begin{array}{ccllccl}
    <r^2> & = & a^2\left(1+\frac{3}{2}e^2\right), & & \displaystyle{\left<\frac{1}{r^3}\right>} & = & \displaystyle{\frac{1}{a^3(1-e^2)^{3/2}}}, \\
    <(\r\cdot\boldsymbol{u})^2> & = & a^2\left(\frac{1}{2}+2e^2\right), & & <(\r\cdot\v)^2> & = & a^2\left(\frac{1}{2}-\frac{1}{2}e^2\right),\\
    \displaystyle{\left<\frac{(\r\cdot\boldsymbol{u})^2}{r^5}\right>} & = & \displaystyle{\left<\frac{(\r\cdot\v)^2}{r^5}\right>} & = \displaystyle{\frac{1}{2a^3(1-e^2)^{3/2}}}.\\
  \end{array}
\end{equation*}
where $<\quad>$ denotes the average over $M$, 
the mean anomaly of the orbit. 

\noindent Let $\displaystyle{\j\equiv\sqrt{1-e^2}\; \n \,\text{,}\;\; \exc = e\boldsymbol{u} \text{,}\;\; \tau=\sqrt{\frac{GM}{a^3}}t \text{,}\;\; \epsp=\frac{J_2R_p^2}{a^2} \;\;\text{and}\;\; \epss=\frac{M_{_{3b}} a^3}{Ma^3_{_{3b}}(1-\es^2)^{3/2}}}$ 
where $\exc$ is the eccentricity vector and $\epsp\geq 0$, $\epss\geq 0$. 
We finally define a dimensionless (divided by $GM/a$) Hamiltonian 
\begin{equation}\label{HamTrem}
  \K'=-\frac{1}{2}+\frac{\epsp}{4(1-e^2)^{5/2}}\Big[1-e^2-3(\j\cdot\np)^2\Big] 
  +\frac{3\epss}{8}\Big[5(\exc\cdot\ns)^2-(\j\cdot\ns)^2-2e^2\Big],
\end{equation}
That describes the secular equations of motion of a test particule 
around an oblate central body perturbed by the third body gravitational 
effect. Let us summarize the used assumptions: 
\begin{enumerate}
  \item the precession rate of the central body spin due to third body 
    gravity is negligible;
  \item the satellite is a massless test particule;
  \item the third body is far enough from the central body such that the 
    third body gravity can be approximated by a quadrupole;
  \item the satellite is far enough from the central body so that the 
    potential from the central body can be approximated as a monopole 
    plus a quadrupole;
  \item the perturbing forces ($\Phi_{_{\!J2}}+\Phi_{_{3b}}$) are weak enough 
    so that the secular equations of motion can be used to describe 
    the orbital motion;
  \item there are not resonant relations in mean motions 
    between the frequencies of the satellite and the frequencies 
    of the central body. 
\end{enumerate}

Let us remark that \citet{SanJuan2006} already studied 
the orbit dynamics about planetary satellites using an 
extensive averaging method based on the Lie transforms 
to obtain averaged equations involving higher orders 
whose result is the introduction of an asymmetry 
for direct and retrograde satellite. Our simplified model will not be 
able to capture this asymmetry because the resulting 
Hamiltonian~(\ref{Hmoyen}) will be symmetric in the satellite 
inclination; thus direct and retrograde satellites will 
have the same behavior.

Let us now make some assumptions suitable in the case of a 
non-inclined central body orbit (e.g. Sun-Mercury-orbiter 
system or Jupiter-Europa-orbiter system). We hereby consider an 
equatorial third body, thus $\np=\ns$. 
We also set $G=\sqrt{1-e^2}$ and $H=G\;\cos \imath$ 
where $\j\cdot\np=\sqrt{1-e^2}\; \cos \imath$. 
To eliminate an extra parameter, we divide the Hamiltonian by 
the coefficient $\epsp$ and we introduce the coefficient $\gamma$ 
\begin{equation*}
  \gamma=\frac{\epss}{\epsp} = 
  \frac{M_{_{3b}}}{Ma^3_{_{3b}}(1-\es^2)^{3/2}}\frac{a^5}{J_2R_p^2}.
\end{equation*}
In Figure~\ref{laplaceSurf}, we represent the geometry 
for the general problem (on the left) and for our simplified one 
(on the right). 
\\
The averaged Hamiltonian is then
\begin{equation}\label{Hmoyen}
  \left\{
  \begin{array}{r}
    \displaystyle{\K'=\frac{\epsp}{4G^3} \left( 1-3\frac{H^2}{G^2} \right)  +  \frac{3\epss}{8} \left[ 5(1-G^2)\left(1-\frac{H^2}{G^2}\right)\sin^2\omega-H^2-2+2G^2\right]}\\
    \displaystyle{\stackrel{{{\K'/\epsp}\atop{\text{ noted}}}\atop{\text{by }\K}}{\iff} \K=\frac{1}{4G^3} \left( 1-3\frac{H^2}{G^2} \right)  +  \frac{3\gamma}{8} \left[ 5(1-G^2)\left(1-\frac{H^2}{G^2}\right)\sin^2\omega-H^2-2+2G^2\right]}.
  \end{array}
  \right.
\end{equation}
This Hamiltonian is 
independent of the ascending node $\Omega$. 
If we take $\gamma=0$ ($\epss=0$, namely we take into account 
only the oblateness effect), we have the well-known circular 
dynamics of the eccentricity vector due to the $J_2$ coefficient 
with an elliptical fixed point in the semi-equinoctial elements 
$(k,h)=(\sqrt{1-G^2}\cos \omega,\sqrt{1-G^2}\sin \omega)$. 
If we take $\gamma\rightarrow\infty$ ($\epsp=0$ i.e. only the third body 
contribution does matter), we find the Kozai-Lidov Hamiltonian which 
we find in a similar formalism in \cite{Paskowitz2004} (with $\es=0$). 
The Hamiltonian (\ref{Hmoyen}) (with $\es=0$) can also 
be found in \cite{Scheeres2001}. 

\begin{table}[h!]
  \begin{center}
    \caption{Connection between semi-major axis (km) of the probe 
      around the central body and the coefficient $\gamma$. 
      The rows ``Min.'',  ``Missions'' and ``Hill'' give the values 
      of $\gamma$ with respect to the semi-major axis 
      respectively equal to the radius of the central body, 
      to one space mission and to the radius of the Hill's sphere.  
      The ``-\;\!-'' symbol indicates that the semi-major axis 
      is lower than the radius of the central body 
      or greater than the radius of the Hill's sphere.}
    \label{lienGamma_Dga_Tab}
    \begin{small}
      \begin{tabular}{lc|>{$}c<{$}>{$}c<{$}>{$}c<{$}|>{$}c<{$}>{$}c<{$}}
        \hline \hline
        & & \text{Min.} & \text{Missions} & \text{Hill} & \multicolumn{2}{c}{\text{Particular values}}\\
        \hline
        \multirow{3}{*}{\green{\text{Mercury}}} & \multirow{2}{*}{$a$ (km)} & \multirow{2}{*}{2\,439.990} & \text{Messenger} & \multirow{2}{*}{175\,295} & \multirow{2}{*}{4\,350} & \multirow{2}{*}{5\,577}\\
        & & & 10\,136.2 & & &\\
        & $\gamma$ & 0.008 & 9.9136 & 1.533\times 10^7 & 1/7 & 0.5\\
        \hline
        \multirow{3}{*}{\yellow{\text{Venus}}} & \multirow{2}{*}{$a$ (km)} & \multirow{2}{*}{6\,051.8} & \text{Venus Express} & \multirow{2}{*}{1\,004\,270} & \multirow{2}{*}{9\,350} & \multirow{2}{*}{12\,010}\\
        & & & 39\,176.8 & & &\\
        & $\gamma$ & 0.008 & 184.485 & 2.042\times 10^9 & 1/7 & 0.5\\
        \hline
        \multirow{3}{*}{\blue{\text{Earth}}} & \multirow{2}{*}{$a$ (km)} & \multirow{2}{*}{6\,378.137} & \text{Meteosat} & \multirow{2}{*}{1\,471\,506} & \multirow{2}{*}{36\,350} & \multirow{2}{*}{46\,670}\\
        & & & 42\,164.14 & & &\\
        & $\gamma$ & 2.38\times 10^{-5} & 0.30107 & 1.559\times 10^7 & 1/7 & 0.5\\
        \hline
        \multirow{3}{*}{\red{\text{Mars}}} & \multirow{2}{*}{$a$ (km)} & \multirow{2}{*}{3\,396.190} & \text{Mars Express} & \multirow{2}{*}{982\,748} & \multirow{2}{*}{26\,150} & \multirow{2}{*}{33\,580}\\
        & & & 9\,311.95 & & &\\
        & $\gamma$ & 5.288\times 10^{-6} & 8.195\times 10^{-4} & 1.073\times 10^7 & 1/7 & 0.5\\
        \hline
        \multirow{3}{*}{\color[rgb]{1,0.5,0}{\text{Europa}}} & \multirow{2}{*}{$a$ (km)} & \multirow{2}{*}{1\,565.0} & \text{EJSM/JEO} & \multirow{2}{*}{13\,529} & \multirow{2}{*}{--} & \multirow{2}{*}{--}\\
        & & & 3\,222 & & &\\
        & $\gamma$ & 1.153 & 42.646 & 5.568\times 10^4 & 1/7 & 0.5
      \end{tabular}
    \end{small}
  \end{center}
\end{table}

\begin{figure}[htbp]
  \begin{center}
    \includegraphics[draft=false,width=1\textwidth]{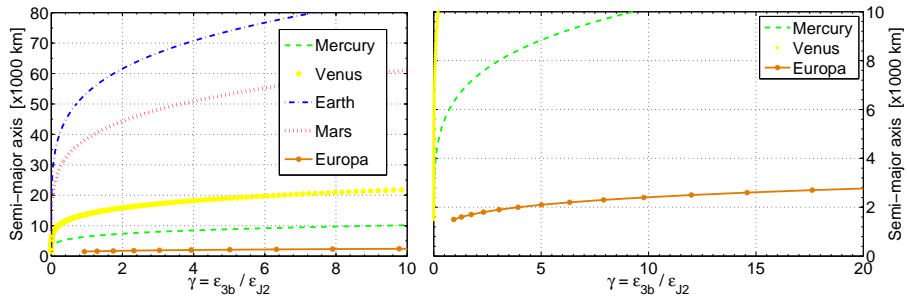}
    \caption{Relation between $\gamma$ and the semi-major axis of 
      a test particule orbiting the central body (terrestrial planets 
      or Europa). The third body are respectively the Sun and Jupiter.}
    \label{lienGamma_Dga}
  \end{center}
\end{figure}

For illustration, we respectively show in the Table~\ref{lienGamma_Dga_Tab} 
and draw in Figure~\ref{lienGamma_Dga} 
the value of the coefficient $\gamma$ with respect to the semi-major axis 
for a probe around a terrestrial planet and around Europa. 
This coefficient can be related to other parameters 
used in the literature. For example, it can be 
linked to the coefficient $\beta$ in \cite{SanJuan2006} or to the 
coefficient $\epsilon$ used in \cite{Scheeres2001}.



\section{Secular Equations of Motion}

From the doubly averaged Hamiltonian (\ref{Hmoyen}), we obtain 
the equations of motion:
\begin{eqnarray}
  \dot{\Omega} & = & -H \Bigg\{\frac{3}{2G^5}+\frac{3\gamma}{8} \left[ \frac{10}{G^2}(1-G^2)\sin^2\omega+2 \right] \Bigg\}\label{OmegaDot}\\
  \dot{H} & = & 0\nonumber\\
  \dot{\omega} & = & \frac{3\gamma}{4} \left[ 5\left(\frac{H^2}{G^3}-G\right)\sin^2\omega+2G \right] + \frac{3}{4G^4}\left( 5\frac{H^2}{G^2}-1 \right) \label{omegaDot}\\
  \dot{G} & = & -\frac{15\gamma}{4}(1-G^2)\left(1-\frac{H^2}{G^2}\right)\sin\omega\cos\omega\;.\label{GDot}
\end{eqnarray}
Developing these equations in eccentricity up to second order, 
we can obtain the equations of \citet{Scheeres2001}. 
In the following, we will adopt a complementary approach, keeping 
functions of eccentricity and inclination, without power 
series developments, in such a way that our results hold for any 
arbitrary eccentricities and inclinations. 

From the previous set of equations, we observe that $H$ is a constant 
of motion, and $H^2=G^2\cos^2\imath$ too, 
as in the 
Kozai-Lidov effect (\citealt{Kozai1962,Lidov1963}). Besides, 
let us remark that $0\leq G \leq 1$, thus 
$0\leq H\leq G\leq 1$ and moreover $\gamma>0$. 
The first equation (\ref{OmegaDot}) is equal 
to zero only for $H=0$ corresponding to exact polar inclination. 
Moreover the ascending node does not affect any of the other 
orbital elements. The last equation (\ref{GDot}) equals to zero 
for $G=1$, $\omega=k\pi/2,\; k\in\N$ or $H=G$, namely $\imath=0\Deg$, 
that is the planar case. The third equation (\ref{omegaDot}) could equal 
to zero for $\omega=0,\pi$ or $\omega=\pm\pi/2$. We analyze these 
equations in next section to find the equilibria.

\section{Frozen Orbit Solutions}
A frozen orbit is characterized by no secular 
change in orbital eccentricity and argument of pericenter. 
It has constant values of $e$, $\imath$ and $\omega$ on average, 
this results in fixed geometrical size and locations, apart from 
short period oscillations.

We already observed that equilibria appear when $G=1$ or 
$\omega=0,\pi$ or $\omega=\pm\pi/2$. We separatly deal with these 
three different cases. For each of them, we give the number of equilibria, 
the conditions of existence and we calculate their stability.

We do not deal with the singularity 
$G=0$ ($\Leftrightarrow e=1$) because 
it corresponds to an escape of the orbiter. 
We will show that the equilibrium $G=1$ always exists. So to begin, 
we deal with the non-circular case $G\neq 1$ (eccentricity $\neq 0$).

\subsection{Non-circular case $G\neq 1$ (eccentricity $\neq 0$)}

\subsubsection{Vertical equilibria -- Kozai-Lidov equilibria: $\cos\omega=0 \Leftrightarrow \omega=\pm\pi/2$}
The conditions to simultaneously equal to zero the equations 
(\ref{omegaDot}) and (\ref{GDot}) is: 
\begin{equation}\label{CondEqPiDemi}
  \left\{
  \begin{array}{rcl}
    \displaystyle{H^2} &=&\displaystyle{\;\frac{G^2}{5}\;\;\frac{1+3G^5\gamma}{1+G^3\gamma}}\\
    \displaystyle{\cos\omega}&=&0
  \end{array}
  \right.
\end{equation}
Because $0\leq G< 1$ then this equation implies that 
\begin{equation}\label{CondStabKozLid}
  H^2<\frac{1+3\gamma}{5\gamma+5}\;.
\end{equation}
Let us observe that this is also the value for which one real 
root does exist. 
If this condition is violated then no real root exists.\\
Actually we determine a region given by the implicit equation 
\begin{equation*}
  \left\{
  \begin{array}{l}
  \displaystyle{864\,000\,H^{16}\gamma^6\,+\,\Big(2\,963\,520\,H^{12}\,-\,1\,024\,H^{10}\Big)\gamma^4}\\
  \displaystyle{\qquad \qquad +\,\Big(1\,512\,630\,H^8\,-\,13\,965\,H^6\,-\,22\,235\,661\,H^{10}\Big)\,\gamma^2\,+\,12\;=\;0}\\
  \text{and }\quad H^2 \leq \frac{1}{3087}
  \end{array}
  \right.
\end{equation*}
where it is possible to find three real roots. We will show that 
these three reals roots appear for eccentricities larger than $0.996\,59$. 
Being a case close to an escape of the orbiter, we will leave to 
section~\ref{LocalDef} a discussion of this ``local deformation''.

If the oblateness term is neglected 
($\epsp\simeq 0 \Leftrightarrow\gamma\rightarrow\infty$), 
the existence condition becomes independent of the physical 
parameter and reduces to $\sin^2 \imath < \frac{2}{5}$ or 
$\arccos\sqrt{\frac{3}{5}}\simeq 39.23\Deg\leq \imath \leq 144.77\Deg$ 
which corresponds to Kozai-Lidov critical inclination.

We also analyze the stability of these equilibria~(\ref{CondEqPiDemi}). 
The Jacobian of the Hamiltonian (\ref{Hmoyen}) evaluated at the equilibrium 
(\ref{CondEqPiDemi}) (noted by $|_{\ateq{CondEqPiDemi}}$ or $G_{kl}$ 
being the value of G at the Kozai-Lidov equilibrium) is given by: 
\begin{equation}\label{S2pidemi}
  \left\{
  \begin{array}{lcl}
    \vspace{0.2cm}
    \displaystyle{\left.\jacc{\K}{G}\right|_{\ateq{CondEqPiDemi}}} &=& \displaystyle{\left.\frac{3}{2G^5}\left(2-15\frac{H^2}{G^2}\right)-\frac{9\gamma}{4}\left( 1+5\frac{H^2}{G^4} \right)\right|_{\ateq{CondEqPiDemi}}}\\
    \vspace{0.2cm}
    &=&\displaystyle{\frac{3}{4 G_{kl}^5} \; \frac{1}{1 + \gamma G_{kl}^3} \Big(-2 + \gamma G_{kl}^3 -21 \gamma G_{kl}^5 -12 \gamma^2 G_{kl}^8\Big)}\\
    \displaystyle{\left.\jacc{\K}{\omega}\right|_{\text{Eq.(\ref{CondEqPiDemi})}}}&=& \displaystyle{\left.-\frac{15}{4}\gamma(1-G^2)\left(1-\frac{H^2}{G^2}\right)\right|_{\text{Eq.(\ref{CondEqPiDemi})}}}\\
    \vspace{0.2cm}
     &=&\displaystyle{-\frac{3}{2}\gamma (1-G^2_{kl})\left(\frac{2-G^5_{kl}\gamma}{1+G^3_{kl}\gamma}\right)}\\
    \vspace{0.2cm}
    \displaystyle{\left.\jac{\K}{G}{\omega}\right|_{\text{Eq.(\ref{CondEqPiDemi})}}}&=&\displaystyle{\left.\jac{\K}{\omega}{G}\right|_{\text{Eq.(\ref{CondEqPiDemi})}}=0}.
  \end{array}
  \right.
\end{equation}
In the equations (\ref{S2pidemi}), the term 
$\left.\jacc{\K}{\omega}\right|_{\text{Eq.(\ref{CondEqPiDemi})}}$ 
is always strictly negative (if $G<1$). Then the equilibrium is a 
stable point if 
\begin{equation}
  \left.\frac{1}{G^5}\left(2-15\frac{H^2}{G^2}\right)-\frac{3\gamma}{2}\left( 1+5\frac{H^2}{G^4} \right)\right|_{\text{Eq.(\ref{CondEqPiDemi})}}<0
  \Longleftrightarrow -2-21\gamma G^5_{kl} +15 \gamma G_{kl} H^2<0\;.\label{S1pidemi}
\end{equation}
This equation (\ref{S1pidemi}) is always satisfied for all 
$\gamma>0$, $H^2<\frac{1+3\gamma}{5\gamma+5}$ and $G_{kl}<1$ ($e_{kl}>0$). 
Therefore, for these conditions, we have two opposite stable points at 
$\omega=\pm \pi/2$ and G such that 
$H^2=\frac{G^2}{5}\;\frac{1+3G^5\gamma}{1+G^3\gamma}$.

\begin{figure}[htbp]
  \begin{center}
    \includegraphics[draft=false,width=1\textwidth]{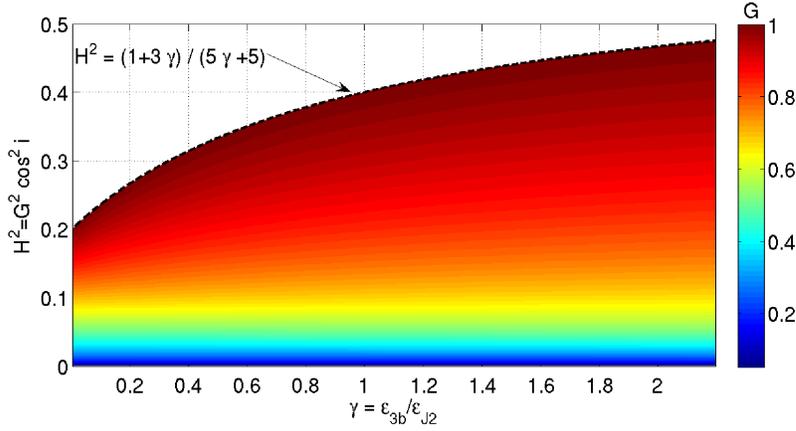}
    \caption{Values of $G$ at Kozai-Lidov stable (Eqs. \ref{CondEqPiDemi} 
      and \ref{CondStabKozLid}) equilibria 
      (vertical equilibria: $\omega=\pm\pi/2$) computed as a 
      function of $H^2$ and $\gamma$. 
      These equilibria are always stable. 
      The color code indicates the value of G at the equilibrium.}
    \label{LocalicationEqPiDemi}
  \end{center}
\end{figure}

For the inclination $\imath=90\Deg$ ($H^2=0$) the equilibrium 
exists for the particular value of $G=0$ ($e=1$). This case is 
only theoretical and should not be considered, because it would 
correspond to an escape of the orbiter. 
In Figure~\ref{LocalicationEqPiDemi} we give the 
location of the equilibria $G$ (\ref{CondEqPiDemi}) in the 
parameter space ($\gamma$,$H^2$).

\subsubsection{Horizontal equilibria: $\sin\omega=0 \Leftrightarrow\omega=0,\pi$}
\begin{figure}[htbp]
  \begin{center}
    \includegraphics[draft=false,width=1.0\textwidth]{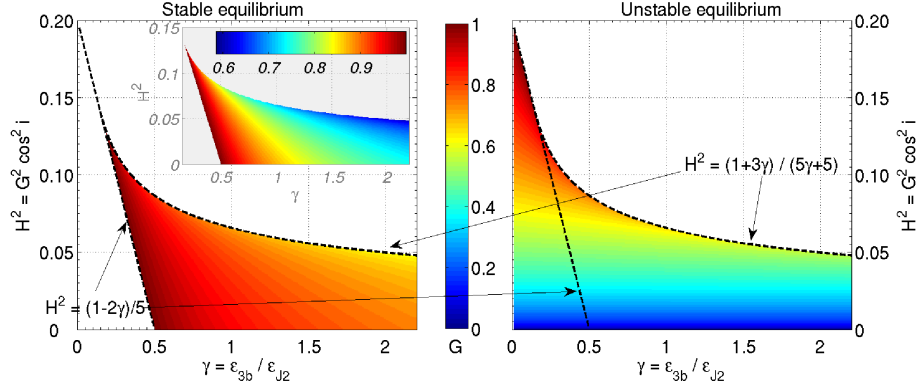}
    \caption{Value of G at stable (on the left)
      (Eqs. \ref{S1Eq0} and condition \ref{CondStabHoriz}) equilibrium 
      and unstable (on the right) (Eqs. \ref{S1Eq0} and condition 
      \ref{CondINStabHoriz}) equilibrium, computed as a function 
      of $H^2$ and $\gamma$.
      The color code indicates the value of G at the 
      equilibrium (horizontal equilibria: $\omega=0,\pi$). 
      The inset on the left panel shows the same plot of left 
      panel but using a wider color code.}
    \label{LocalicationEqHoriz}
  \end{center}
\end{figure}

The conditions to simultaneously equal to zero the equations 
(\ref{omegaDot}) and (\ref{GDot}) are: 
\begin{equation}\label{CondEq0}
  \left\{
  \begin{array}{rcl}
    \displaystyle{H^2} &=&\displaystyle{\;\frac{G^2}{5}\;\;(1-2G^5\gamma)}\\
    \displaystyle{\sin\omega}&=&0.
  \end{array}
  \right.
\end{equation}
Using {\it``Le théorème d'algèbre de Sturm''} \citep{Sturm} 
(for more explanation see the Appendix) we calculate the number 
of roots ($G$) in the range $0\leq G<1$ of the equation 
(\ref{CondEq0}) as a function of the parameters $\gamma$ and $H^2$. 
For $\gamma>0$ this equation has 
\begin{itemize}
  \item[$\bullet$] one real root, equal to 0 
    if $H^2=0$ and $\gamma< 1/2$.
  \item[$\bullet$] three real roots (one equal to 0 
    and the other two opposite) if $H^2=0$ and $\gamma\geq 1/2$.
  \item[$\bullet$] three real roots (one equal to 1 
    and the other two opposite) if $0<H^2<\frac{1-2\gamma}{5}$;
  \item[$\bullet$] five real roots (one equal to 1 
    and the other ones opposite two by two) if $\gamma\geq 1/7$ and 
    $\frac{1-2\gamma}{5}<H^2<\frac{(7\gamma)^{-2/5}}{7}$;
  \item[$\bullet$] one real root equal to 1 
    otherwise.
\end{itemize}
In Figure~\ref{LocalicationEqHoriz}, we give the 
location of the equilibria $G$ (\ref{CondEq0}) in the space ($\gamma$,$H^2$). 
The particular case $G=1$ will be treated in the next section. 
We can also analyze the stability of these equilibria~(\ref{CondEq0}). 
The Jacobian of the Hamiltonian (\ref{Hmoyen}) evaluated at the equilibrium 
(\ref{CondEq0}) (noted by $|_{\ateq{CondEq0}}$ or $G_{hor}$, being 
$G_{hor}$ the value of G at the equilibrium) is given by: 
\begin{equation}\label{S1Eq0}
  \left\{
  \begin{array}{lcl}
    \vspace{0.2cm}
    \displaystyle{\left.\jacc{\K}{G}\right|_{\ateq{CondEq0}}} &=& \displaystyle{\left.\frac{3}{2G^5}\left(2-15\frac{H^2}{G^2}\right)+\frac{3\gamma}{2}\right|_{\ateq{CondEq0}}}\\
    \vspace{0.2cm}
    &=&\displaystyle{\frac{3}{2G^5_{hor}}\Big(-1+7\gamma G^5_{hor}\Big)}\\
    \vspace{0.2cm}
    \displaystyle{\left.\jacc{\K}{\omega}\right|_{\ateq{CondEq0}}}&=& \displaystyle{\left.\frac{15}{4}\gamma(1-G^2)\left(1-\frac{H^2}{G^2}\right)\right|_{\ateq{CondEq0}}}\\
    &=&\displaystyle{\frac{3}{2}\gamma (1-G_{hor}^2) (2+G_{hor}^5\gamma)}\\
    \vspace{0.2cm}
    \displaystyle{\left.\jac{\K}{G}{\omega}\right|_{\ateq{CondEq0}}}&=&\displaystyle{\left.\jac{\K}{\omega}{G}\right|_{\ateq{CondEq0}}=0}.
  \end{array}
  \right.
\end{equation}
In the equations (\ref{S1Eq0}), the term 
$\left.\jacc{\K}{\omega}\right|_{\ateq{CondEq0}}$ 
is always strictly positive (if $G<1$). Then the equilibrium is a 
stable point if 
\begin{equation*}
  \gamma>\frac{1}{G^5}\left(15\frac{H^2}{G^2}-2\right) \Longleftrightarrow 
  G^5_{hor}>\frac{1}{7\gamma}.
\end{equation*}
Using this equation at the equilibrium (\ref{CondEq0}), we obtain 
conditions for stability of the stable point ($G\neq 1\Leftrightarrow e\neq 0$)
\begin{equation}\label{CondStabHoriz}
  \left\{
  \begin{array}{rcl}
    \displaystyle{\frac{1-2\gamma}{5}} & < & H^2 < \displaystyle{\frac{1}{7} \left(\frac{1}{7\gamma}\right)^{2/5}}\\
    \text{and}\quad\displaystyle{\frac{1}{7}} & \leq & \gamma\,.
  \end{array}
  \right.
\end{equation}
So the condition to have an unstable equilibrium is given by
\begin{equation}\label{CondINStabHoriz}
  \gamma < \frac{1}{7} \quad\text{ or }\quad H^2 < \frac{1-2\gamma}{5}\quad
  \text{ or }\quad H^2 > \frac{1}{7} \left(\frac{1}{7\gamma}\right)^{2/5}\,.
\end{equation}

In the Figure~\ref{LocalicationEqHoriz}, we notice that when both 
unstable and stable equilibrium exist, the unstable equilibrium 
always appears for a value of G lower than that of the stable point  
(i.e. for a value of $e$ greater than the one for the stable point).

\subsection{Circular case $G=1$ (eccentricity $e=0$)}
For the case $G=1$, we can use a canonical transformation to 
cartesian coordinates 
\begin{equation}
  x=\sqrt{1-G^2}\sin\omega\qquad \qquad y=\sqrt{1-G^2}\cos\omega
\end{equation}
The new Hamiltonian is therefore 
\begin{eqnarray}
  \K&=&\frac{1}{4}\left(\frac{1}{(1-x^2-y^2)^{3/2}}-\frac{3H^2}{(1-x^2-y^2)^{5/2}} \right) \nonumber\\
  \label{hamxy}\\
  &&+\;\frac{3\gamma}{8} \left[ 5x^2\left(1-\frac{H^2}{1-x^2-y^2}\right)-H^2-2x^2-2y^2 \right]\nonumber
\end{eqnarray}
for which it is obvious that $(0,0)$ is always an equilibrium point 
whose stability can be studied computing the second derivatives 
and evaluate them at this equilibrium: 

\begin{equation*}
  \left\{
  \begin{array}{lcl}
    \vspace{0.3cm}
    \displaystyle{\jacc{\K}{x}\Big|_{x=0=y}} & = & \displaystyle{\frac{3}{4}(1-5H^2)+\frac{3\gamma}{4}(3-5H^2)}\\
    \displaystyle{\jacc{\K}{y}\Big|_{x=0=y}} & = & \displaystyle{\frac{3}{4}(1-5H^2)-\frac{3\gamma}{2}}\\
    \displaystyle{\jac{\K}{x}{y}\Big|_{x=0=y}} & = & \displaystyle{\jac{\K}{y}{x}\Big|_{x=0=y}\; =\;0\,.}
  \end{array}
  \right.
\end{equation*}
So, the condition to have a stability point at $x=0=y$ is
\begin{equation}\label{CondStab00}
  H^2<\frac{1-2\gamma}{5}\qquad \text{ or }\qquad H^2>\frac{1+3\gamma}{5\gamma+5}\,;
\end{equation}
and thus the condition to have an unstable point at $x=0=y$ is 
\begin{equation}\label{CondINStab00}
  \frac{1-2\gamma}{5} < H^2 < \frac{1+3\gamma}{5\gamma+5}\,.
\end{equation}

\subsection{Summary of the phase space}
\begin{figure}[htbp]
  \begin{center}
    \hspace{-1cm}
    \includegraphics[draft=false,width=1.07\textwidth]{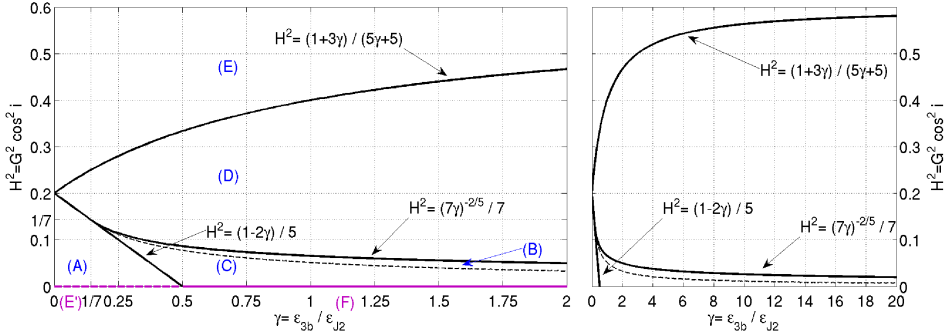}
    \caption{Bifurcation lines and regions in the parameter 
      space ($\gamma$,$H^2$). In the regions (B) and (C) we have the same 
      number and the same stability of the equilibria but the phase 
      space is topologically different. These regions are separated 
      by the dashed line implicitely given by 
      equation~(\ref{CondSepEspPhBC}).}
    \label{PhSp}
  \end{center}
\end{figure}

In this section we summarize the various possible phase spaces topologies 
as a function of the parameters. 
We draw (Fig.~\ref{PhSp}) the bifurcation lines 
(conditions \ref{CondStabKozLid}, \ref{CondStabHoriz} and \ref{CondStab00})
in the parameter space ($\gamma$,$H^2$). This bifurcation 
diagram is equivalent to the upper part of the bifurcation 
diagram in \cite{SanJuan2006} but here we draw the bifurcation lines 
in the general (not linked to a particular central body) 
space ($\gamma$,$H^2$). The $H^2=(7\gamma)^{-2/5}/7$ line stops at the 
limit $\gamma=1/7$. For this value, this curve coincides 
with the $H^2=(1-2\gamma)/5$ condition. 
For the Jupiter-Europa-orbiter system, the minimum value of $\gamma$ 
is $1.153$ (Tab.~\ref{lienGamma_Dga_Tab}). Therefore the phase spaces 
(A) and (E') do not exist.

The region (E') and (F) in magenta color 
correspond to exact polar orbits ($\imath=90\Deg$ thus $H^2=0$). 

\begin{figure}[htbp]
  \vspace{1cm}
  \begin{center}
    \hspace{-0.92cm}
    \begin{minipage}{0.336\textwidth}
      \includegraphics[draft=false,width=1.16\textwidth]{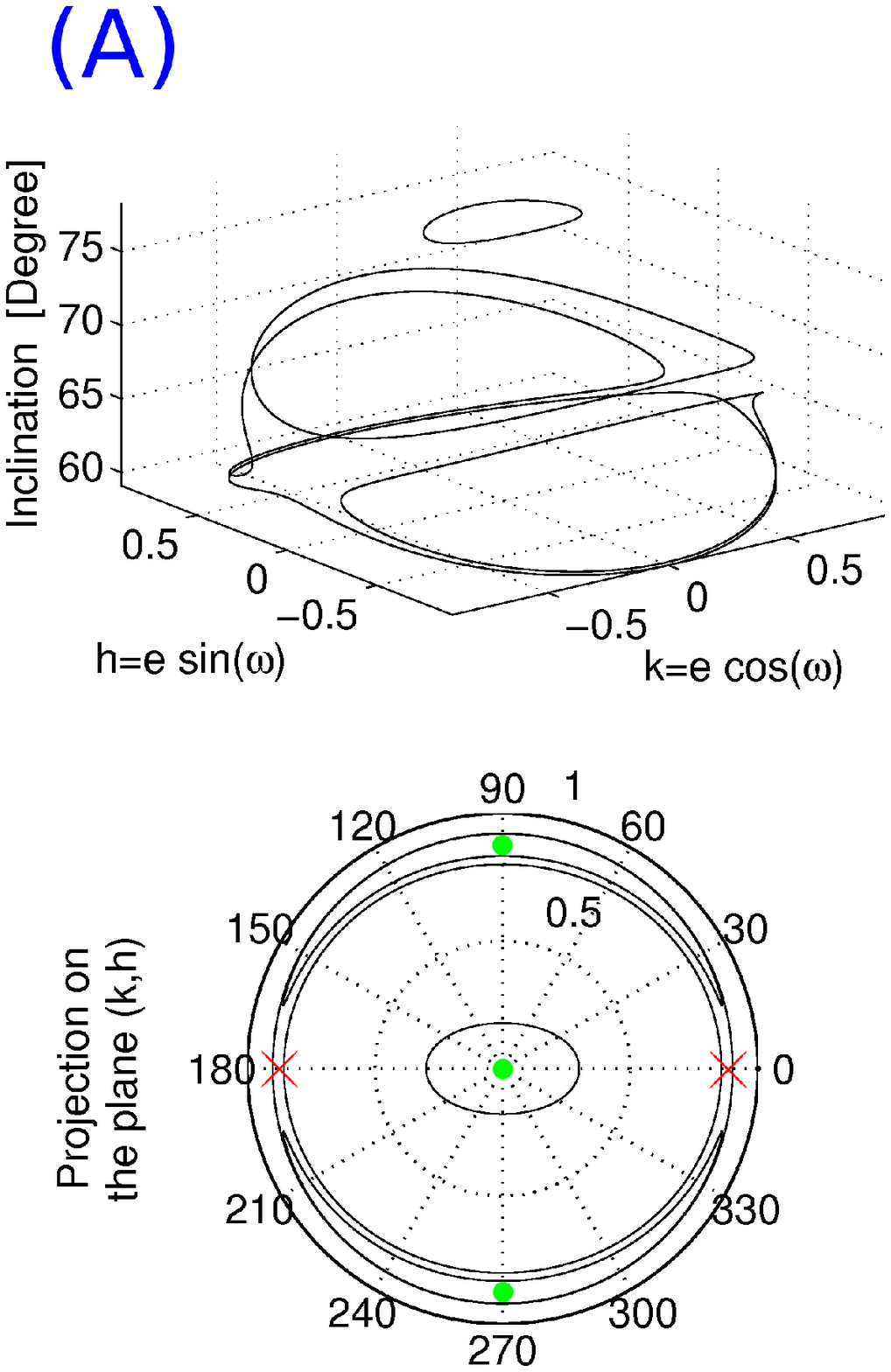}
    \end{minipage}
    \hspace{0.1cm}
    \begin{minipage}{0.336\textwidth}
      \includegraphics[draft=false,width=1.16\textwidth]{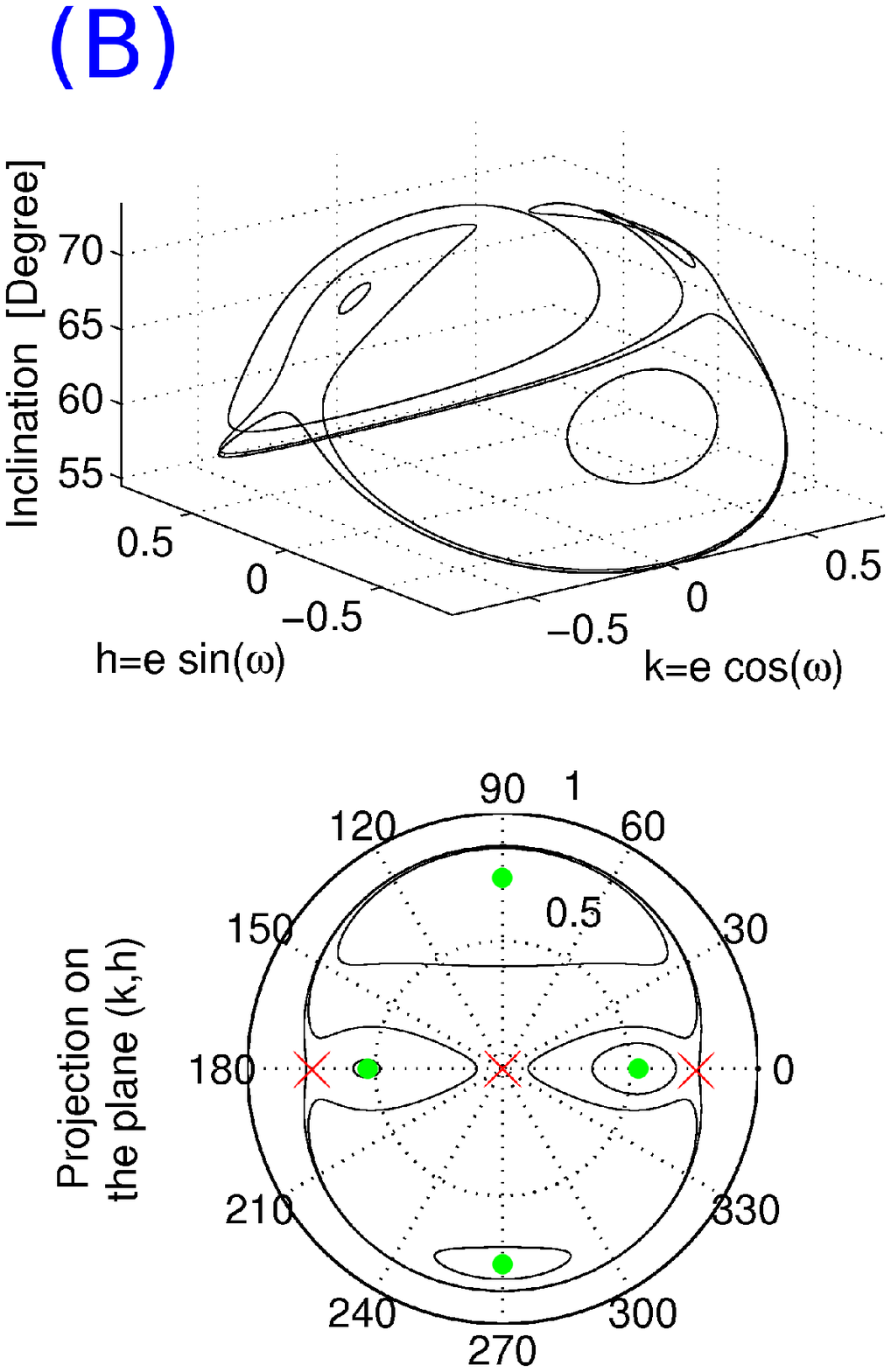}
    \end{minipage}
    \hspace{0.1cm}
    \begin{minipage}{0.336\textwidth}
      \includegraphics[draft=false,width=1.16\textwidth]{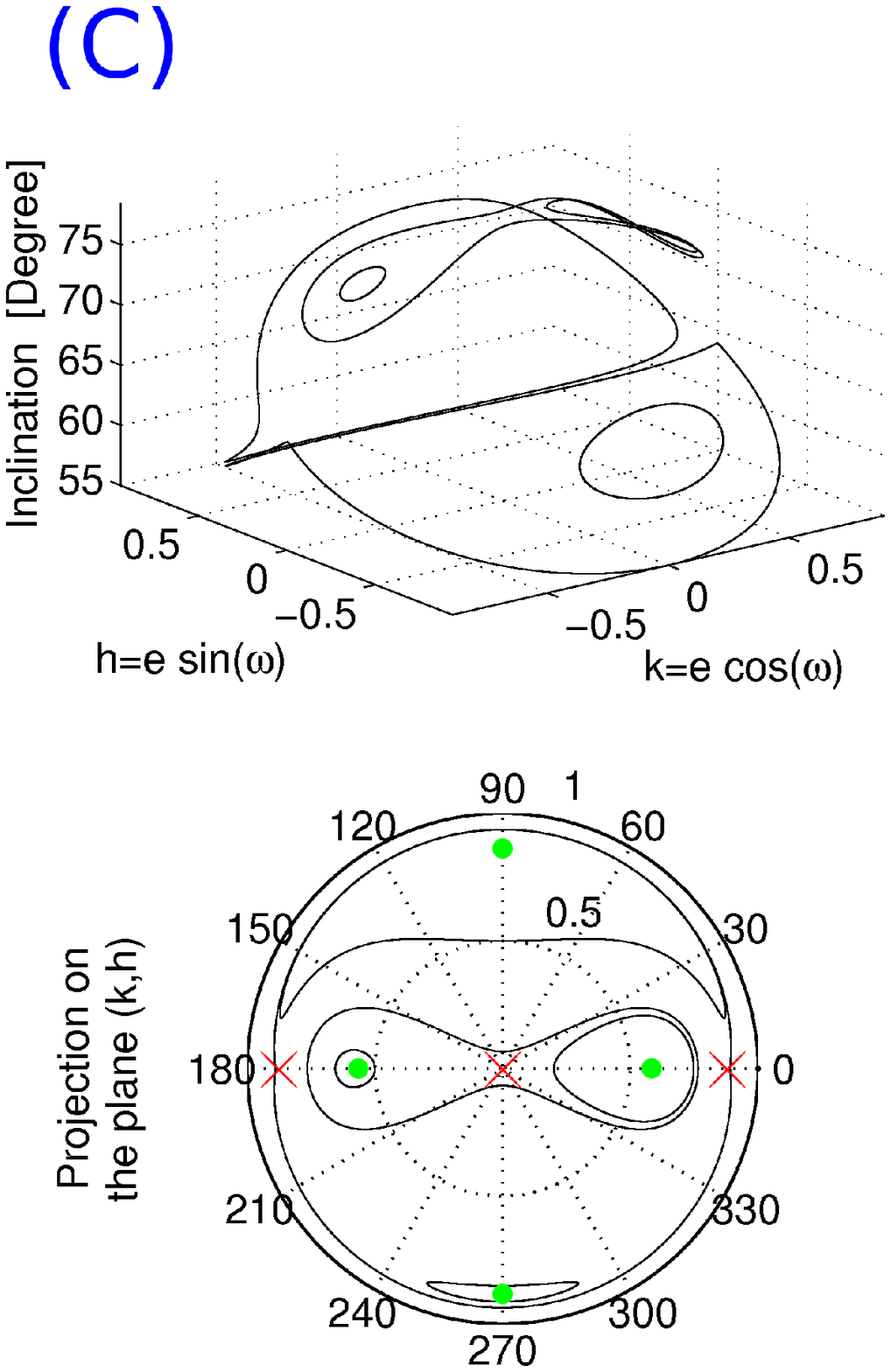}
    \end{minipage}\\

    \vspace{0.5cm}
    \rule{\linewidth}{.5pt}
    \noindent\hrulefill
    \vspace{0.5cm}

    \hspace{-0.92cm}
    \begin{minipage}{0.336\textwidth}
      \includegraphics[draft=false,width=1.16\textwidth]{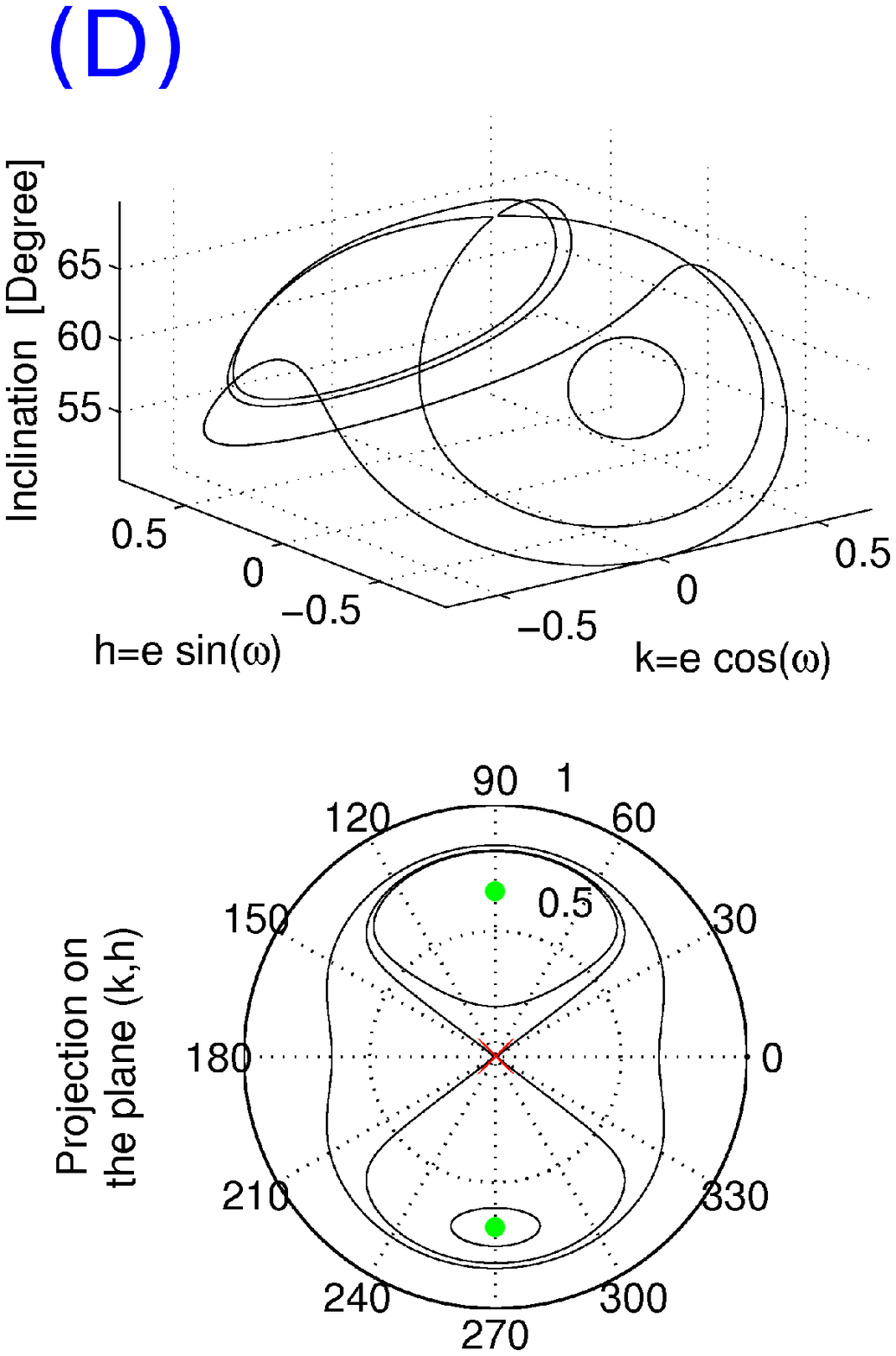}
    \end{minipage}
    \hspace{0.1cm}
    \begin{minipage}{0.336\textwidth}
      \includegraphics[draft=false,width=1.16\textwidth]{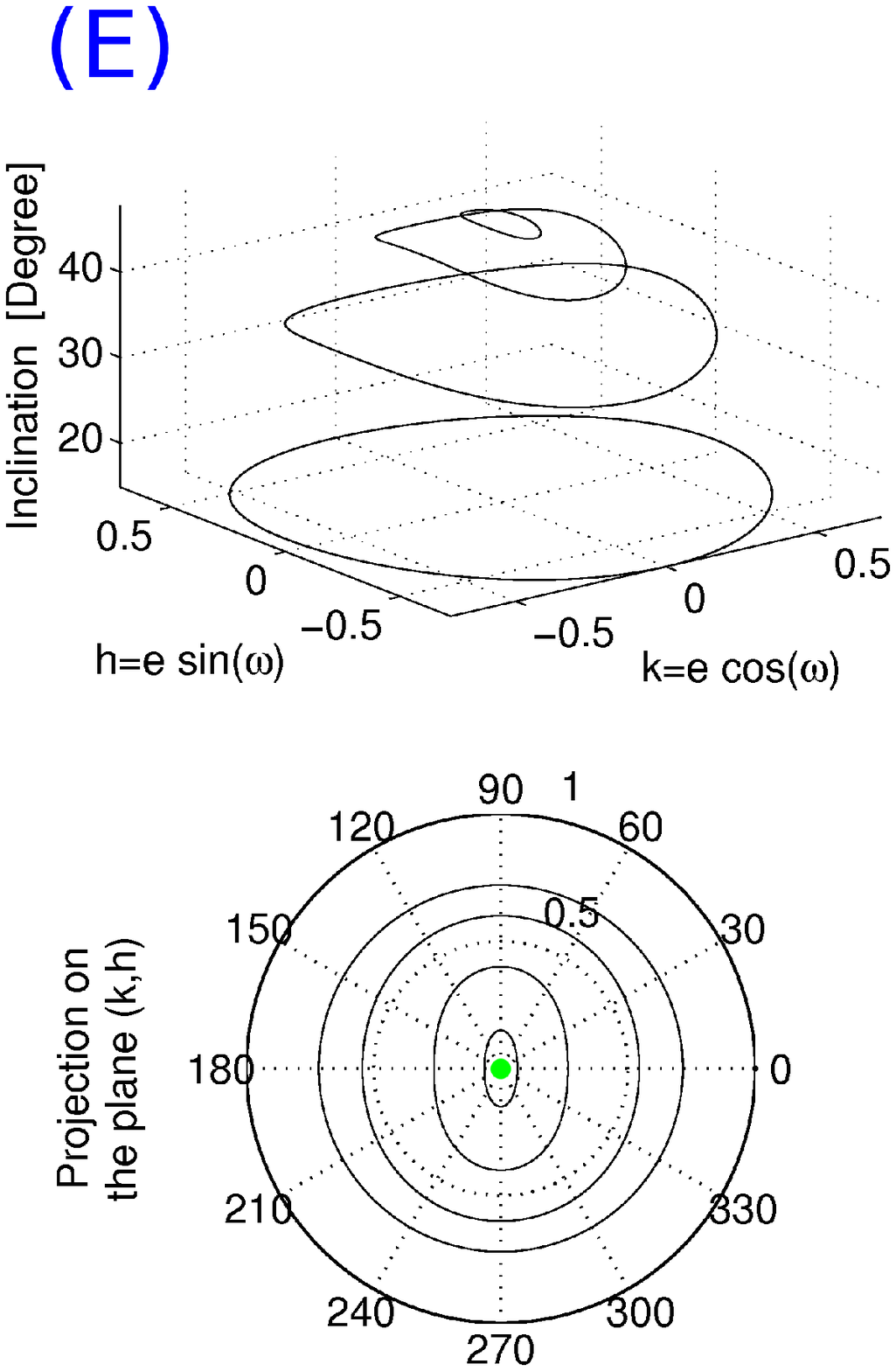}
    \end{minipage}
    \hspace{0.1cm}
    \begin{minipage}{0.336\textwidth}
      \includegraphics[draft=false,width=1.16\textwidth]{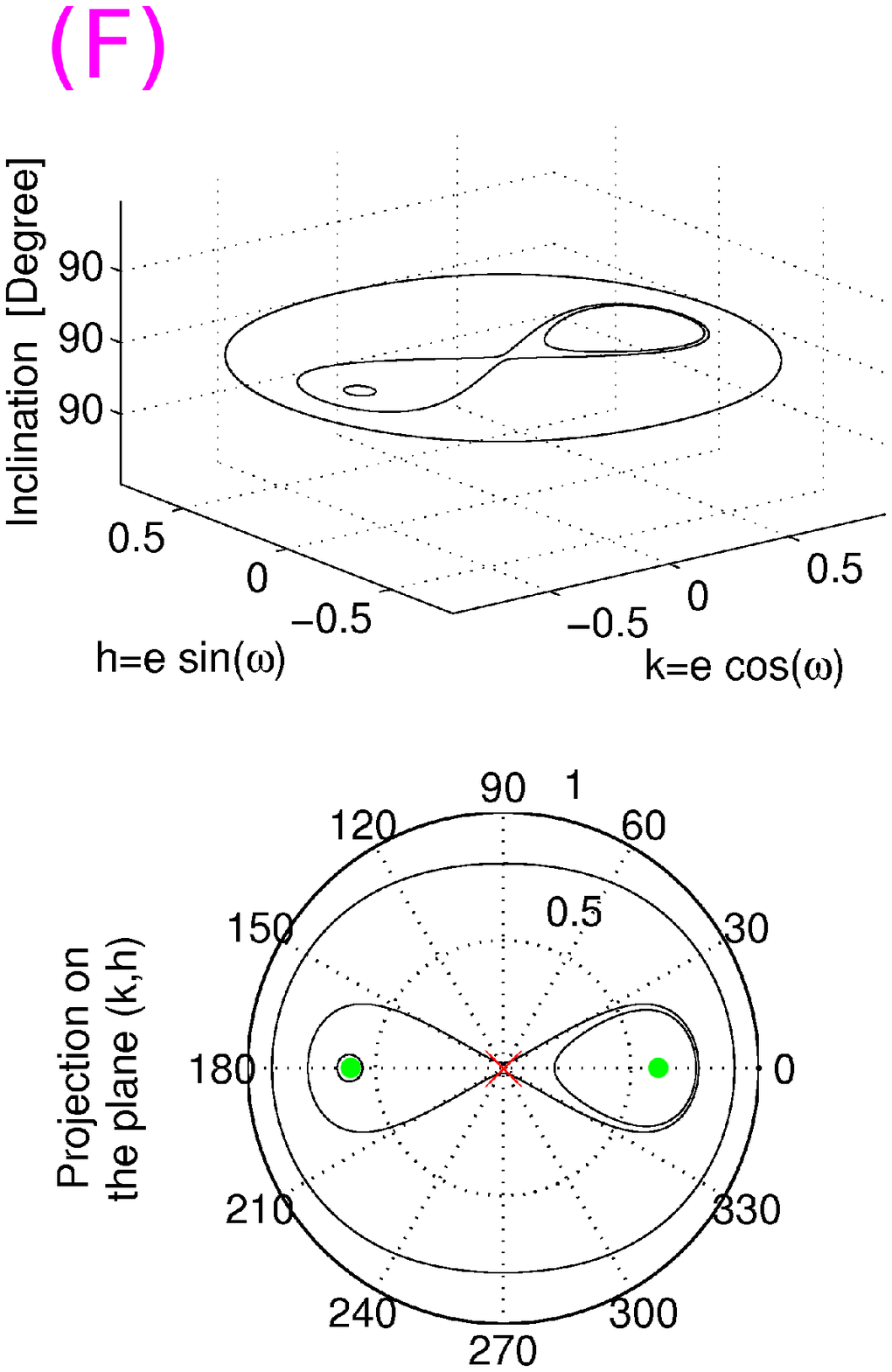}
    \end{minipage}
    \caption{Examples of some generic contour plots of the Hamiltonian 
      (\ref{Hmoyen}) in $(k,h,\imath)$ space for each region of 
      Fig.~\ref{PhSp}. The inclination $\imath$ is given in degrees 
      and the semi-equinoctial elements $(k,h)$ are given by 
      $(k,h)=(\sqrt{1-G^2}\cos \omega,\sqrt{1-G^2}\sin \omega)$. 
      The green point and red cross are respectively the stable 
      and unstable points. In the polar projection, the radius 
      is the eccentricity $e$ and the angle is the 
      pericenter $\omega$ in degrees. }
    \label{PhSpDC}
  \end{center}
\end{figure}

For each region, we attribute a letter and we draw (Fig.~\ref{PhSpDC}) 
a generic contour plot of the Hamiltonian (\ref{Hmoyen}) in the 
$(k,h,\imath)$ physical space. We recall that 
the motion of the inclination~$\imath$ is given by the conservation 
of the first integral $H=G\;\cos \imath$. We also draw the 
projection of these phase spaces in the semi-equinoctial elements space 
$(k,h)=(\sqrt{1-G^2}\cos \omega,\sqrt{1-G^2}\sin \omega)$. 
In this phase space, it is easier to bring to the fore the 
stable (green point) and unstable (red cross) equilibria. 
The (E') phase space is trivial, containing only concentric circle in 
the $\imath=90\Deg$ plane, so we do not repoduce it. 

In  the Figure~(\ref{PhSpDC}), we notice that the 
maximum inclination is always reached at $e=0$. 
This is explained by the relation $H^2=\sqrt{1-e^2}\cos\imath$. 
This last relation also gives a maximum bound onto the eccentricity: 
$e\leq \sqrt{1-H^2}$. Therefore there are some values of $H$ 
for which the phase space is visibly restricted in eccentricity. 
Beyond this eccentricity, the motion is physically impossible.

Let us observe that the region near the stable equilibria allows to 
control the variation of the eccentricity even for high eccentricity. 
We also remark that there are ``dangerous'' portions of phase space 
such as the region around the $\gamma=H^2=1/7$ or near of 
the (B)-(C) transition. In these regions the dynamics 
(in a full model) could change strongly 
for a small variation of ($H^2,\gamma$) or ($e,\omega$).

The transition between (B) and (C) phase spaces arises when the 
energy of the separatrix at the $(0,0)$ equilibrium is equal to the energy 
of the unstable exterior horizontal equilibrium. This condition gives 
a new ``fictitious'' bifurcation line (dashed line in Figure~\ref{PhSp}) 
in the parameter space $(\gamma,H^2)$. To find this line, we evaluate the 
Hamiltonian (\ref{Hmoyen}) at the unstable equilibrium 
$H^2=\frac{G^2}{5}(1-2G^5\gamma)$ 
(Eq.\ref{CondEq0} and condition \ref{CondINStabHoriz}) and we 
denote this value by $\K_{1}$. Afterward, we evaluate the Hamiltonian 
(\ref{hamxy}) at the unstable equilibrium $(0,0)$ 
(condition \ref{CondINStab00}) and denote the result by $\K_{2}$. 
We now assume these two equilibria have the same value of Hamiltonian $\K$
and of $H^2$. Then we can replace $H^2$ by $\frac{G^2}{5}(1-2G^5\gamma)$ 
in $\K_2$ and we impose the equality between $\K_1$ and $\K_2$. 
Therefore we obtain the condition 
\begin{equation}\label{CondSepEspPhBC}
  \gamma = \frac{2+5G_{hor}^3+3G_{hor}^5}{6G_{hor}^{10}+15G_{hor}^3-21G_{hor}^5}\quad
  \text{ where } H^2=\frac{G_{hor}^2}{5}(1-2G_{hor}^5\gamma)
\end{equation}
where $G_{hor}$ is the unstable horizontal equilibrium i.e. 
$\displaystyle{H^2=\frac{G_{hor}^2}{5}(1-2G_{hor}^5\gamma)}$. 
We plot this implicit condition (\ref{CondSepEspPhBC}) in 
Figure~\ref{PhSp} with a dashed black line. This line joins the 
``$(1-2\gamma)/5$'' and ``$(7\gamma)^{-2/5}/7$'' lines 
at the $(\gamma=1/7,H^2=1/7)$ point.

For the particular case $\gamma \rightarrow 0$ ($J_2$ effect only), 
we obtain, for all $H^2$, a phase space with circular motion 
of the eccentricity. We see that near to the value $H^2=1/2$ 
(corresponding to the Molniya\footnote{At this inclination, 
due to $J_2$ effect, the argument of perigee remains nearly 
constant for a long period of time. Molniya orbits are named 
after a series of Soviet/Russian Molniya communications 
satellites which have been using this type of orbit 
since the mid 1960s.} critical inclination equal to 
$ 63.43\Deg$ with $G=1$), the phase spaces (A), (D) and (E) 
always exist until $\gamma$ becomes exactly equal to $0$. 
In the opposite case, $\gamma\rightarrow\infty$ (third body effect 
only), the curve $H^2=(7\gamma)^{-2/5}/7$ converges 
to $0$ and the curve $H^2=(1+3\gamma)/(5\gamma+5)$ converges to $3/5$ 
(corresponding to the Kozai-Lidov critical inclination equal 
to $39.23\Deg$ with $G=1$). 
Then only the following phase spaces are realizable: 
(E) (for $H^2>3/5$), (D) (for $0<H^2<3/5$) and (F) (for $H^2=0$) 
with (F) that degenerates to an unstable point at the 
center. These three phase spaces will be shown in 
Figure~\ref{HamKLEspPh}.

\begin{figure}[htbp]
  \begin{center}
    \includegraphics[draft=false,width=1.0\textwidth]{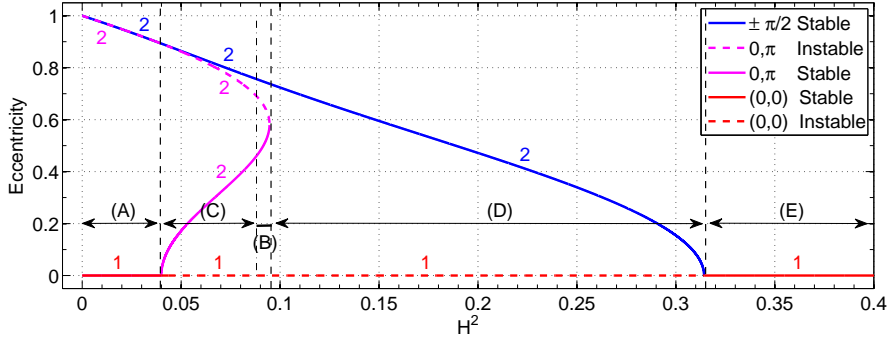}
    \caption{For a vertical section ($\gamma=0.4$) in the 
      Figure~\ref{PhSp}, value of the eccentricity at unstable 
      (dashed color lines) and stable (solid color lines) equilibria 
      with respect to $H^2$. 
      The numbers give the number of equilibria for each curve.}
    \label{BifLineGammaCst}
  \end{center}
\end{figure}
In Figure~\ref{BifLineGammaCst}, we show how the stable and unstable 
equilibria evolve, appear and disappear in each region and during the 
transition between the regions. We take a vertical section in the 
Figure~\ref{PhSp} at $\gamma=0.4$. This section crosses the regions 
(A), (C), (B), (D) and (E). We draw the value of the eccentricity 
at the stable (solid color lines) and unstable (dashed color lines) 
equilibria with respect to $H^2$. The vertical dashed black 
lines mark the boundary of the regions. The numbers give the number 
of equilibria with this value of $e$. For example, $2$ in magenta 
dashed line means that there are two unstable equilibria with the 
same value of $e$, respectively for $\omega=0$ and $\omega=\pi$.

At the transition between (A) and (C), the central ($e=0$) stable 
point bifurcates in two horizontal stable points 
($e\neq 0$ and $\omega=0,\pi$) and one unstable point ($e=0$). 
At the transition between (B) and (D), the two unstable and the two 
stable horizontal ($\omega=0,\pi$) equilibria converge to the same 
value of $e$ and cancel out. 
At the transition between (D) and (E), the two stable Kozai-Lidov 
equilibria ($\omega=\pm \pi/2$) come close to 0 and cancel out 
with the central unstable equilibrium to give one central stable 
equilibrium. We remark that the transition between (C) and (B) is 
not characterized by a change of the equilibria.

\subsection{Period at the equilibrium}
We are now interested in the period of the eccentricity vector 
at the equilibrium. This will be done by linearizing in a 
neighborhood of the equilibrium. Then the Hamiltonian close 
to the equilibrium is given by (the subscript $eq.$ 
means ``evaluated at the equilibrium''): 
\begin{eqnarray*}
  \K&=&\K_{eq.} \;+\; \underbrace{\dpar{\K}{G}\Big|_{eq.}}_{=0}\;(G-G_{eq.}) \;+\;
  \underbrace{\dpar{\K}{\omega}\Big|_{eq.}}_{=0}\;(\omega-\omega_{eq.}) \\
  &\;+&\underbrace{\frac{1}{2}\jacc{\K}{G}\Big|_{eq.}}_{\stackrel{not.}{=}a}\;(\underbrace{G-G_{eq.}}_{\stackrel{not.}{=}X})^2  \;+\;
  \underbrace{\jac{\K}{G}{\omega}\Big|_{eq.}}_{=0}\;(G-G_{eq.})(\omega-\omega_{eq.})  \;+\;
  \underbrace{\frac{1}{2}\jacc{\K}{\omega}\Big|_{eq.}}_{\stackrel{not.}{=}b}\;(\underbrace{\omega-\omega_{eq.}}_{\stackrel{not.}{=}Y})^2\\
  \K&=&\K_{eq.}+aX^2+bY^2\;.
\end{eqnarray*}
This is an harmonic oscillator that can be expressed in action-angle 
variables ($\psi, J$) defined as (at a stable equilibrium, we have $ab>0$):
\begin{equation*}
  X=\sqrt[4]{\frac{b}{a}}\sqrt{2J}\cos\psi \qquad \text{ and } 
  \qquad  Y=\sqrt[4]{\frac{a}{b}}\sqrt{2J}\sin\psi\,.
\end{equation*}
Then the frequency at the equilibrium is given by 
\begin{equation}\label{FreqEq}
  \dot{\psi}=\dpar{\K}{J}=2\sqrt{ab}=\sqrt{\jacc{\K}{G}\Big|_{eq.}\jacc{\K}{\omega}\Big|_{eq.}}\;.
\end{equation}
Using the equation (\ref{FreqEq}), the periods ($\tau$) 
at the stable equilibria are given by:
\begin{itemize}
\item for horizontal equilibria: $G$ such as Equation~(\ref{CondEq0}) 
  and condition of stability~(\ref{CondStabHoriz})
  \begin{equation*}
    \tau = \frac{4\pi} { 3 \sqrt{\frac{5}{2}  
        \left[ \frac{\epsp}{G^5} \Big(2-15\frac{H^2}{G^2}\Big)+\epss \right]
        \epss(1-G^2)\left(1-\frac{H^2}{G^2}\right) }}\,;
  \end{equation*}
\item for vertical (Kozai-Lidov) equilibria: $G$ such as 
  Equation~(\ref{CondEqPiDemi}) and condition of 
  stability~(\ref{CondStabKozLid})
  \begin{equation*}
    \tau = \frac{4\pi}{3\sqrt{-\frac{5}{2} 
        \left[ \frac{\epsp}{G^5}\Big(2-15\frac{H^2}{G^2}\Big) - 
          \frac{3\epss}{2}\left(1+5\frac{H^2}{G^4}\right) \right] 
        \epss(1-G^2)\left(1-\frac{H^2}{G^2}\right)} }\,;
  \end{equation*}
\item for central equilibrium ($e=0$): $G=1$ with condition of 
  stability~(\ref{CondStab00})
  \begin{equation*}
    \tau = \frac{8\pi}{3\sqrt{ \Big[\epsp(1-5H^2)+\epss(3-5H^2)\Big]  
        \Big[\epsp(1-5H^2)-2\epss\Big]}}\,.
  \end{equation*}
\end{itemize}
We remind that $\gamma=\epss/\epsp$ and that the equations are 
dimensionless. Then the periods at the equilibria are given 
by $T_{eq.}=\sqrt{\frac{a^3}{GM}}\,\tau_{eq.}$. 

\begin{figure}[htbp]
  \begin{center}
    \includegraphics[draft=false,width=1.0\textwidth]{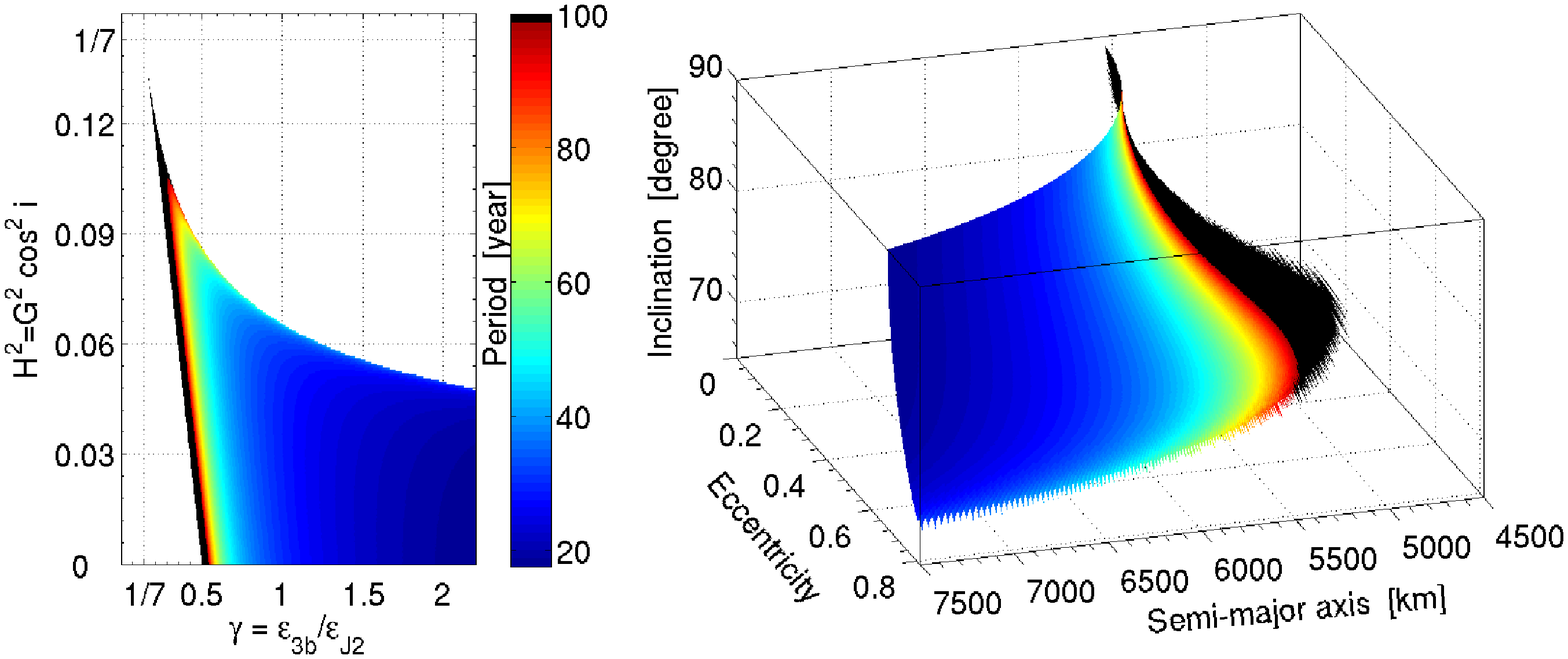}
    \includegraphics[draft=false,width=1.0\textwidth]{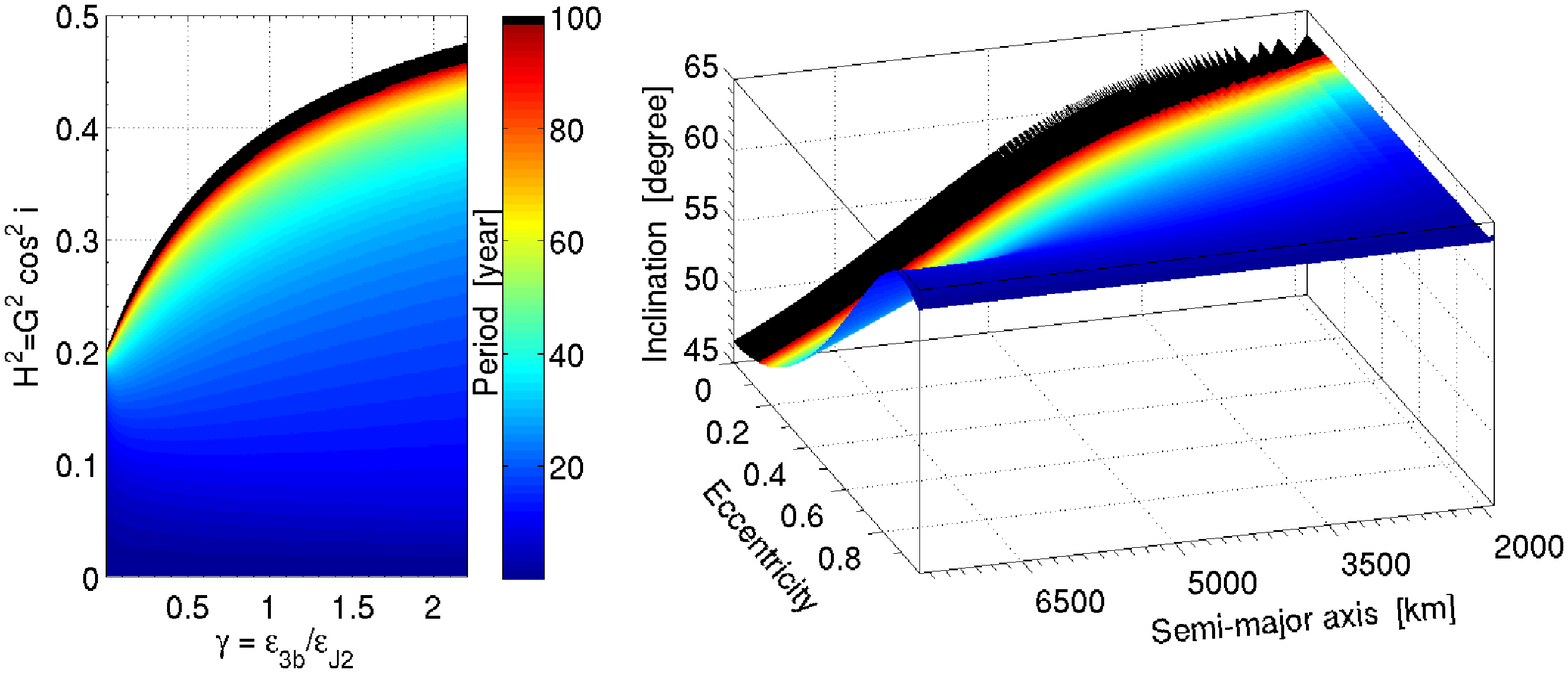}
    \includegraphics[draft=false,width=1.0\textwidth]{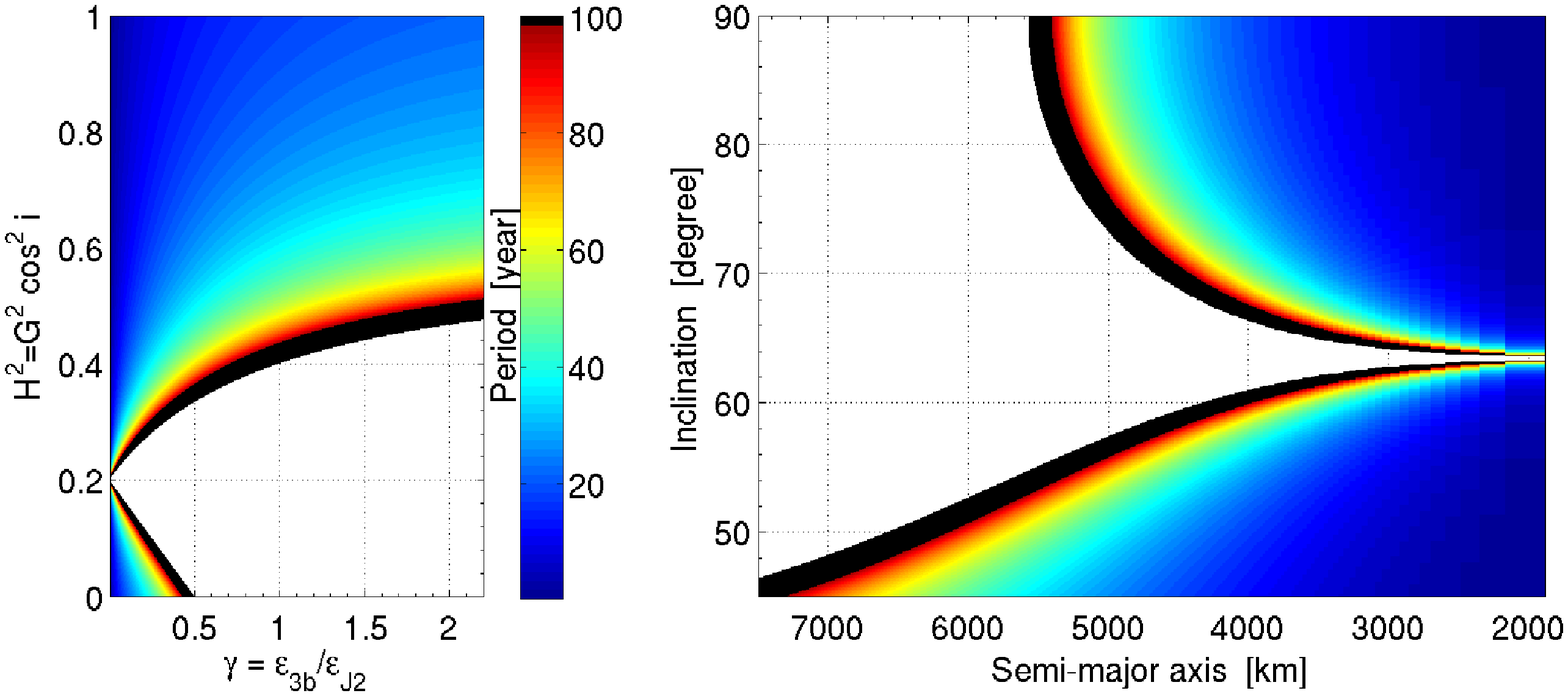}
    \caption{Plot of the periods at the stable horizontal equilibrium 
      (Eq.~\ref{CondEq0} with conditions~\ref{CondStabHoriz}), 
      vertical equilibrium (Eq.~\ref{CondEqPiDemi} with 
      condition~\ref{CondStabKozLid}) and $(0,0)$ equilibrium 
      (with conditions~\ref{CondStab00}) 
      respectively in the upper, center and lower panels. 
      The color code indicates the period (truncated to $100$ years) of 
      the fundamental frequency at the equilibrium. On the left panels, 
      the period with respect to the parameters $(\gamma,H^2)$. On the 
      right panels, the location of the stable equilibrium in the 
      physical space $(a,e,\imath)$ with its period. 
      For the equilibrium $(0,0)$, $e$ is always equal to $0$.}
    \label{PeriodEq}
  \end{center}
\end{figure}

For example, we apply these formula to a Mercury orbiter. 
The values for Mercury are $a_{{_{3b}}}=57\,909\,176.0$ km, 
$\es=0.205\,630\,69$, $J_2=6.0\times 10^{-5}$ \citep{Anderson1987} 
and $R_p=2\,439.99$ km. In the Figure~\ref{PeriodEq} we plot 
the periods at the equilibria respectively for the three cases:
\begin{itemize}
\item on the left panel, the periods at the stable equilibrium with respect to 
  the value of $\gamma$ and $H^2$. The color code indicates the period 
  of the fundamental frequency at the equilibrium;
\item on the right panel, the location of the stable equilibrium in the 
  phase space $(a,e,\imath)$ with the period in the color scale.
\end{itemize}
The color code is the same for the left and right panels and it 
is truncated at the value of $100$ years. For a larger 
period, we use the black color.

\section{Comparison of analytical and numerical solutions}

\subsection{Comparison for all inclinations}
\begin{figure}[htbp] 
  \includegraphics[draft=false,width=1.0\textwidth]{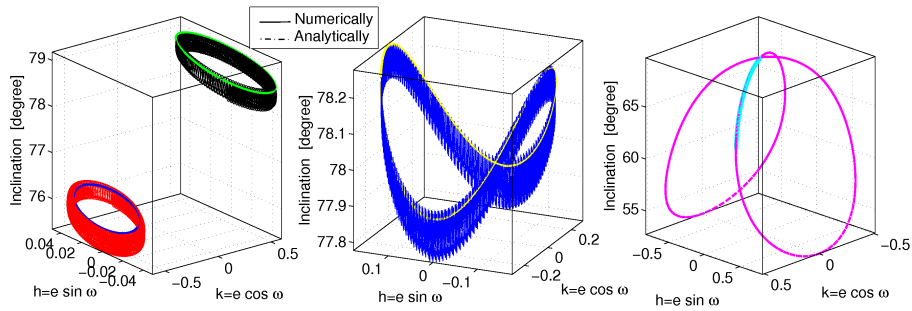}
    \caption{Comparison between analytical and numerical results. 
      For the left and right panel, the initial conditions are 
      $a_0=6\,407$~km ($\gamma=1.000296$), $\Omega_0=M_0=0\Deg$.
      In the left panel, for the lower orbit, we take 
      $e_0=0.545055$ ($G=0.8384$), $\imath_0=76.646989\Deg$ 
      ($H^2=3.7492\times 10^{-2}$) and $\omega_0=180\Deg$; 
      for the upper orbit, we take $e_0=0.6$ ($G=0.8$), 
      $\imath_0=78.221768\Deg$ ($H^2=0.26666666$) and 
      $\omega_0=0\Deg$. For the right panel, the initial conditions 
      are $e_0=0.01$ ($G=0.99995$), $\imath_0=69.73104\Deg$ 
      ($H^2=0.1199999$) and $\omega_0=0\Deg$. 
      For the middle panel, the initial conditions 
      are $a_0=4\,650$~km ($\gamma=0.17096$), $e_0=0.3$ ($G=0.9539392$), 
      $\imath_0=77.89775\Deg$ ($H^2=0.04$) and 
      $\Omega_0=M_0=\omega_0=0\Deg$. 
      The numerical model takes into account the contribution of 
      $J_2$ and $C_{22}$ and the solar gravitational effect, 
      with starting epoch fixed at 14 September 2019. 
      The analytical model is based on Equations (\ref{OmegaDot}, 
      \ref{omegaDot}, \ref{GDot}). 
      We plot the numerical integrations with continued lines and 
      the analytical results with dashed lines. 
      In the right panel, the numerical integration leads to a crash 
      onto the planet.
      \label{CompNumAnal}}
\end{figure}

The analytical results of the simplified model described above 
are checked using a precise numerical integration of the complete 
set of equations of motion (\ref{diffEq}). For our test, we use 
Mercury's orbiter mission profile, which nominally puts the 
spacecraft into a high eccentric polar orbit. Numerical 
integrations were performed with the Bulirsch-Stoer 
\citep{bulirsh-stoer} integrator. We reproduce hereby afew 
characteristic plots of the numerical simulations to confirm 
our analytical theory (see Figure~\ref{CompNumAnal}). 
Similar results have been obtained for a wide range of 
initial frozen orbit conditions. 

Figure~\ref{CompNumAnal} shows a very good agreement between 
analytical results and numerical simulations.

\subsection{Comparison for polar inclination and explanation of the preliminary numerical results}
\begin{figure}[htbp] 
  \includegraphics[draft=false,width=1.0\textwidth]{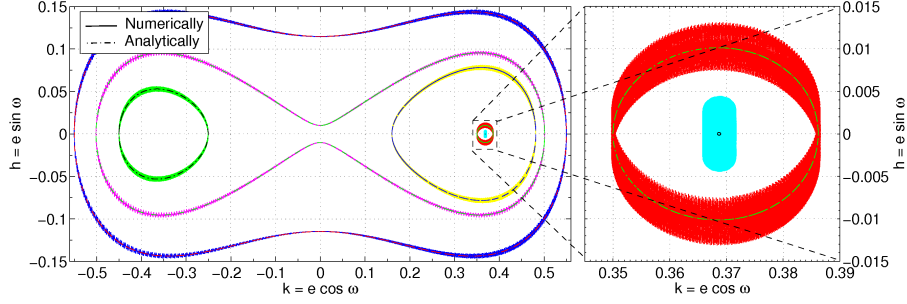}
    \caption{Comparison between analytical and numerical results 
      for exact polar orbiter. The initial conditions are $a_0=6\,000$~km
      ($\gamma \simeq 0.72$), $\imath_0=90\Deg$ , $\Omega_0=67.7\Deg$, 
      $\omega_0=-2\Deg$, $M=36.4\Deg$.
      The numerical and analytical model are the same 
      of Figure~\ref{CompNumAnal}. In dashed line the analytical 
      result and in continued line the numerical integration. 
      On the right a blow-up of the center of libration.
      \label{CompNumAnalI90}}
\end{figure}

In the Figure~\ref{CompNumAnalI90}, we present a graphical comparison 
between numerical integration and analytical results (contour plots of the 
Hamiltonian~(\ref{Hmoyen})) for an exact polar inclination. 
We see that the analytical theory is very close to the numerical 
integration for all initial eccentricities. We also notice that 
the addition of the $C_{22}$ does not modify much the motion.

In the right panel, we show two solutions close to the libration point 
and we see that, the closer the motion is to the libration equilibrium, 
the more the numerical integrations show a discrepancy with respect to 
the analytical results for the periherm libration: 
the frozen orbit of the analytical model shows no changes in 
eccentricity and argument of pericenter. 
On the contrary, the numerical orbit has short period oscillations 
but constant mean values of $e$ and $\omega$. 

Figure~\ref{CompNumAnalI90} allows us to explain the behaviors already 
seen in our preliminary numerical exploration (Fig.~\ref{excMap}). 
In fact we can find there different orbits with a semi-major axis 
equal to $6\,000$ km corresponding to a vertical section in 
Figure~\ref{excMap}. Then, on this section, 
we take some values of the eccentricity such that: 
\begin{itemize}
  \item for $e$ near to $0$, in Fig.~\ref{excMap}, we see a large 
    value of the amplitude of variation of the eccentricity 
    approximatively equal to $0.5$ and a high value of the second 
    derivative.\\
    In Fig.~\ref{CompNumAnalI90}, for $e$ equal to $0$, we are on the 
    separatrix. Therefore the eccentricity increases (roughly until $0.5$)  
    and a little shift of the initial eccentricity causes a high 
    difference of the frequency. Thus the second derivative of the 
    frequency is large;
  \item for $e$ close to $0.37$, in Fig.~\ref{excMap}, we see that the 
    amplitude of variation of the eccentricity decreases until $0$.\\
    In Fig.~\ref{CompNumAnalI90}, at $e=0.37$, we find the stable point 
    where the eccentricity is equal to a constant;
  \item  when $e$ moves away from $0.37$ to $0.5$, in Fig.~\ref{excMap}, 
    we see that the amplitude of variation of the eccentricity 
    increases from $0$ to $0.5$ and for $e=0.5$, the amplitude of 
    variation of the eccentricity is maximal and the value of 
    the second derivative is large.\\
    In Fig.~\ref{CompNumAnalI90}, moving away from the equilibrium 
    ($e=0.37$) toward the separatrix ($e\simeq 0.5$) 
    we encounter larger and larger variations in $e$; 
  \item  for $e$ near to $0.58$, in Fig.~\ref{excMap}, we see that 
    the amplitude of variation of the eccentricity is smaller than 
    for $e\simeq 0.5$.\\
    In Fig.~\ref{CompNumAnalI90}, for $e\simeq 0.58$, the pericenter 
    circulates and the maximum of the amplitude of variation of 
    the eccentricity is roughly equal to $0.58-0.12=0.46$.
  \item in Fig.~\ref{excMap}, moving along the line $e=0$, we pass 
    from the region (F) to the region (E') at $5\,577$ km 
    (Tab.\ref{lienGamma_Dga_Tab}). For semi-major axis smaller 
    than $a=5\,577$ km, we do not cross any separatrix and the 
    amplitude of variation of the eccentricity is small.
\end{itemize}
\subsection{Frequency comparison}

To obtain a second independent validation of our analytical model, 
we numerically compute, using the \texttt{NAFF} algorithm 
\citep{Laskar1988,Laskar2005}, the period of the numerical solutions of 
the full system~(\ref{diffEq}) obtained through numerical integration, 
and we compare it with the period of the equilibrium points of 
the simplified model. 

Table~\ref{ComPerAnalNum} provides a summary of this comparisons 
We can observe a very good agreement between the two methods.
Some small differences can be explained as follows: 
\begin{itemize}
\item the exact equilibrium in the doubly averaged system is not the exact 
  equilibrium in the full numerical model;
\item the full numerical model contains short period terms which 
  disturb the long period dynamics.
\end{itemize}
\begin{table}
  \begin{center}
    \caption{Comparison between the period of  the equilibria 
      determined in the analytical model and the period 
      numerically obtained using \texttt{NAFF}.}
    \label{ComPerAnalNum}
    \begin{tabular}{c|c|c|c|c|c|c}
      \hline \hline
      \multicolumn{4}{c|}{Initial condition}& \multicolumn{2}{c|}{Period  [year]}&Error\\
      \hline
      What & $a$ & $e$ & $\imath$ & Analytical & Numerical & relative\\
      Equi.& [km]&&[degree]&&&\%\\
      \hline
      Kozai & $5\,750$ & $0.4731$ & $58.37$ & $29.30$ & $29.25$ & $0.17$\\
      Horiz.& $8\,083$ & $0.4922$ & $77.68$ & $35.67$ & $35.61$ & $0.17$\\
      Horiz.& $5\,818$ & $0.5418$ & $71.93$ & $42.17$ & $42.26$ & $0.21$ \\
      (0,0) & $3\,429$ & $0.0$ & $47.64$ & $9.127$  & $9.135$  & $0.08$ \\
      (0,0) & $4\,731$ & $0.0$ & $77.01$ & $56.594$ & $55.274$ & $2.38$ \\
      \hline
    \end{tabular}
  \end{center}
\end{table}

\section{Discussions}

\subsection{$J_2$: the protector}
\begin{figure}[htbp]
  \begin{center}
    \includegraphics[draft=false,width=1\textwidth]{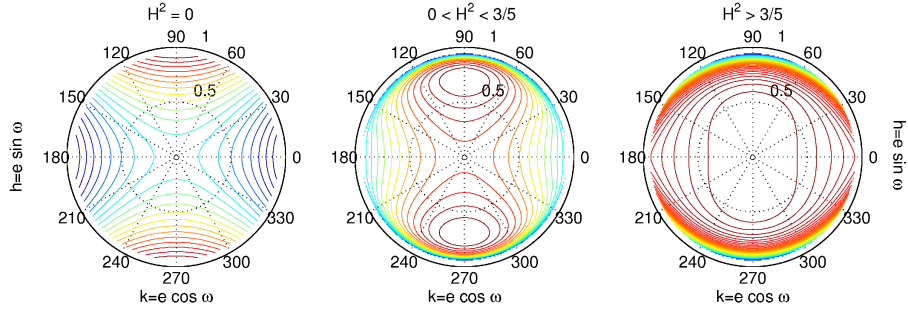}
    \caption{All possible phase spaces for Kozai-Lidov 
      Hamiltonian~(\ref{KozLidHam}) with respect to the values of $H^2$.}
    \label{HamKLEspPh}
  \end{center}
\end{figure}

The aim of this section is to describe the protection mechanism 
of the coefficient $J_2$ on the increase of the eccentricity. 
We recall that our Hamiltonian (\ref{Hmoyen}), once we set the 
coefficient $\epsp=0$, reduces to the Kozai-Lidov Hamiltonian: 
\begin{equation}\label{KozLidHam}
    \K_{kl}=\frac{3\,\epss}{8} \left[ 5(1-G^2)\left(1-\frac{H^2}{G^2}\right)\sin^2\omega-H^2-2+2G^2\right]\,.
\end{equation}
In the Figure~\ref{HamKLEspPh}, we draw the possible phase spaces 
of this Hamiltonian. In the right panel ($H^2>3/5$) we have 
a similar behavior of our (E) case (Fig.~\ref{PhSpDC}). For the exact 
polar orbits ($H^2=0$ in the left panel of the Fig.~\ref{HamKLEspPh}), 
in the Kozai-Lidov Hamiltonian, all the probes are ejected: 
the eccentricity always grows up to $1$. Instead, with the addition 
of the coefficient $J_2$ we have the phase space (E') or (F) 
(Fig.~\ref{PhSpDC}) where it is possible that the eccentricity 
does not increase or that it remains at a fixed value. 
In the middle case ($0<H^2<3/5$) we see that 
for an initial pericenter close to $0$, the eccentricity 
increases. Instead, in our case, the phase spaces (A), (B), (C) and (E) 
(Fig.~\ref{PhSpDC}) show that it is possible to find initial 
condition (other than $\omega\simeq \pm \pi/2$) where the 
increasing of the eccentricity is naturally controlled.

{\it The $J_2$ acts as a protection mechanism against the increase of the eccentricity due to the Kozai-Lidov effect.} 
This mechanism also appears for planets in tight binary systems 
\citep{Saleh2009}, where the general relativistic effects become 
dominant and can cause the periastron to precess on very short 
timescales. Therefore this precession can lead to the suppression 
of Kozai oscillations. 

\subsection{Local deformation of the Kozai-Lidov equilibrium}

\begin{figure}[htbp]
  \begin{center}
    \hspace{-1.5cm}
    \includegraphics[draft=false,width=1.15\textwidth]{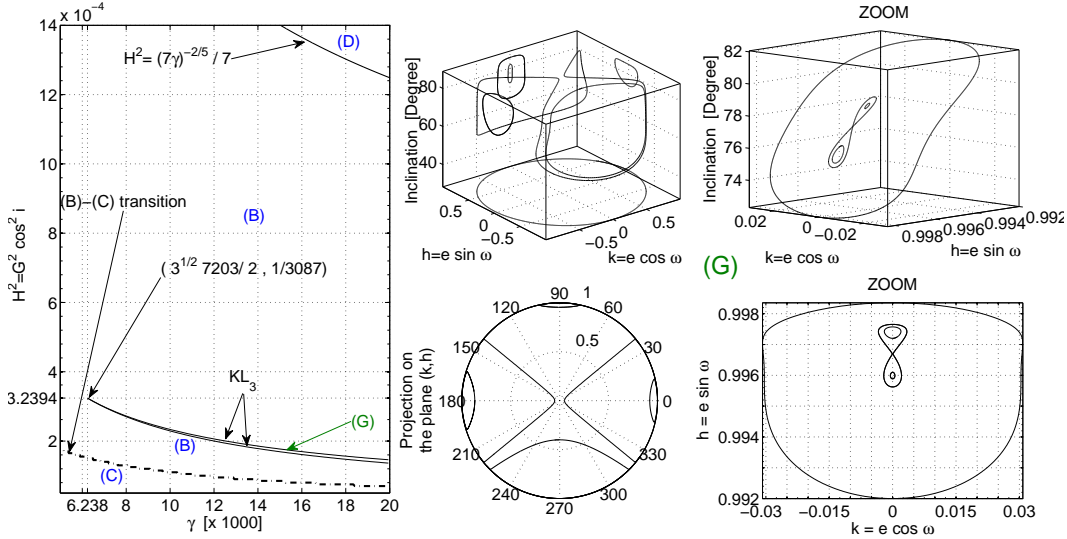}
    \caption{Local deformation of the Kozail-Lidov equilibrium. 
    The bifurcation lines in the left panel with the new region (G) 
    demarcated by the two curves $KL_3$. 
    Example of generic contour (for (G) region) of the Hamiltonian 
    (\ref{Hmoyen}) in $(k,h,\imath)$ space in the middle panels. 
    A zoom of the local deformation in the right panels.}
    \label{LocaDefKL_fig}
  \end{center}
\end{figure}

We have seen that the condition to get the Kozai-Lidov equilibrium is 
(Eq.~\ref{CondStabKozLid})
\begin{equation*}
  H^2<\frac{1+3\gamma}{5\gamma+5}\;.
\end{equation*}
Actually there is a region where it is possible to find three 
real roots for $G$ on a fonction of $H^2$ and $\gamma$. 
The conditions to have these three real roots are given by:
\begin{equation*}
  KL_3\equiv \left\{
  \begin{array}{l}
  \displaystyle{864\,000\,H^{16}\gamma^6\,+\,\Big(2\,963\,520\,H^{12}\,-\,1\,024\,H^{10}\Big)\gamma^4}\\
  \displaystyle{\qquad \qquad +\,\Big(1\,512\,630\,H^8\,-\,13\,965\,H^6\,-\,22\,235\,661\,H^{10}\Big)\,\gamma^2\,+\,12\;=\;0}\\
  \text{and }\quad H^2 \leq \frac{1}{3087}
  \end{array}
  \right.
\end{equation*}
We draw the solutions of this equation, denoted by $KL_3$, 
that demarcates the region denoted (G), on the left panel of the 
Figure~\ref{LocaDefKL_fig}. Let us observe that this condition verified 
for large value of $\gamma$ 
($\gamma\geq 7203\sqrt{3}/2$) 
and for very small value of $H^2$ ($H^2\leq 1/3087$). An example of 
the phase space is plot in Figure~\ref{LocaDefKL_fig} in the middle panels. 
In this region, the vertical Kozai-Lidov stable equilibrium 
bifurcates in two stable and one unstable 
vertical Kozai-Lidov equilibria producing thus a local deformation 
of the Kozai-Lidov equilibrium. We show an example of these three 
equilibria in the right panels of the Figure~\ref{LocaDefKL_fig}. 
Initial conditions close to these equilibria (external orbit in 
the right panels of Fig.~\ref{LocaDefKL_fig}) give rise to orbit 
librating around this set of three equilibria.

It is possible to find that this bifurcation appears, in the (G) region, 
for a value of $G$ smaller than $\sqrt{3}/21\simeq0.082\,478\,6$ 
corresponding to a value of the eccentricity $e$ larger than 
$\sqrt{438}/21\simeq 0.996\,59$. Recalling the formula $H=G\cos\imath$, 
we obtain a minimal inclination of $87.27\Deg$. 

\subsection{$J_3$ discussion}

In \citet{Paskowitz2006} the authors included the $J_3$ 
(the ``{\it pear shape}'' of the central body) Europa's effect 
in their system. They noticed that the coefficient $J_3$ 
caused an asymmetry between the solutions of the frozen 
orbits for $\omega=\pm \pi/2$ but they did 
not explain the reasons of this beavior. 

The potential arising from a central body with a $J_3\neq 0$ is given by 
\begin{equation*}
  \Phi_{_{\!J3}}(\r) = \frac{GM J_3 R_p^3}{2r^7} \;(\r\cdot\np)
  \Big[5(\r\cdot\np)^2-3r^2\Big].
\end{equation*}
The averaged Hamiltonian is then
\begin{equation*}
  \frac{3\;GM J_3 R_p^3}{2\;a^4 (1-e^2)^{5/2}}\; e\; \sin \omega \;\sin \imath 
    \;\left( 1- \frac{5}{4}\sin^2\imath \right).
\end{equation*}
Using our variables $G=\sqrt{1-e^2}$, $H=G\cos\imath$, we can define 
the dimensionless (divided by $GM/a$) potential that we can add 
to the Hamiltonian (\ref{Hmoyen}): 
\begin{equation*}
  \underbrace{\frac{J_3 R_p^3}{a^3}}_{\stackrel{not.}{=}\epsJ3} \; \frac{3}{8G^8}
  \sqrt{1-G^2} \; \sin \omega \; \sqrt{G^2-H^2} \; (5H^2-G^2).
\end{equation*}
Introducing the coefficient 
$\displaystyle{\delta=\frac{\epsJ3}{\epsp}=\frac{J_3 R_p}{J_2 a}}$, 
the equations of motion (\ref{GDot} and \ref{omegaDot}) can be 
rewritten in compact form as follows: 
\begin{equation*}
  \left\{
    \begin{array}{lcl}
      \dot{G} & = & F_1(G,H,\gamma)\; \sin\omega \cos\omega\; +\; F_2(G,H,\delta)\; \cos \omega\\
      \dot{\omega} & = & F_3(G,H,\gamma)\; +\; F_4(G,H,\gamma)\;\sin^2\omega\; +\; F_5(G,H,\delta)\; \sin\omega
    \end{array}
    \right.
\end{equation*}
where the functions $F_1$, $F_3$ and $F_4$ can be easily identified 
in equations (\ref{omegaDot}) and (\ref{GDot}). The 
functions $F_2$ and $F_5$ come from the $J_3$ effect 
and they are proportional to $\delta$.

\subsubsection{Vertical equilibria -- Kozai-Lidov equilibria: $\cos\omega=0 \Leftrightarrow \omega=\pm\pi/2$}
Let us observe that the addition of $J_3$ effect causes an asymmetry 
in the frozen orbit solutions not present before. Indeed, 
for $\omega=\pi/2$ the condition of equilibrium is given by
\begin{equation*}
  F_3(G,H,\gamma) + F_4(G,H,\gamma) + F_5(G,H,\delta) = 0
\end{equation*}
whereas for $\omega=-\pi/2$ the condition of equilibrium is given by
\begin{equation*}
  F_3(G,H,\gamma) + F_4(G,H,\gamma) - F_5(G,H,\delta) = 0.
\end{equation*}
Then, for a small coefficient $\delta$, the asymmetry is not 
important. However for a large value of this coefficient, 
the asymmetry could be important until the elimination 
of one of two equilibria. 

\subsubsection{Horizontal equilibria.}
For horizontal equilibria, the condition of equilibrium 
($\dot{G}=0$) becomes: 
\begin{equation*}
  F_1(G,H,\gamma) \sin\omega + F_2(G,H,\delta) = 0 \iff 
  \sin\omega=-F_2/F_1\stackrel{not.}{=}-\epsilon.
\end{equation*}
Then the ``horizontal'' equilibria appear for non-zero values of the 
pericenter $\omega=-\epsilon$ and $\omega=\pi+\epsilon$. The 
condition to obtain $\dot{\omega}=0$ becomes: 
\begin{equation*}
  F_3(G,H,\gamma) + F_4(G,H,\gamma)\epsilon^2 - F_5(G,H,\delta)\epsilon=0,
\end{equation*}
that induces a shift in the equilibrium in $G$ and $\omega$ variables 
with respect to the case ``$J_2$ + third body''.

\subsubsection{Modifications of the phase space}\label{LocalDef}
\begin{figure}[htbp] 
  \hspace{-2.8cm}
  \includegraphics[draft=false,width=1.4\textwidth]{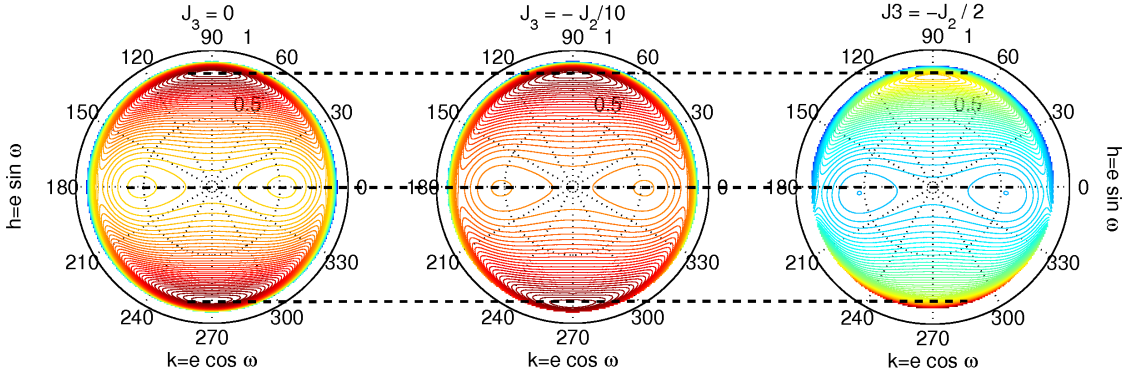}
  \caption{Distortion of the phase space (for a Mercury's orbiter)
    due to $J_3$ effect. The initial conditions are $a=5\;900$ 
    km ($\gamma\simeq 0.66$) and $H^2=0.06$. $J_3$ 
    (respectively $\delta$) is equal to $0$ ($0$), 
    $-J_2/10$ ($-0.041356$) and $-J_2/2$ ($-0.20678$) 
    in left, center and right panels.
    \label{J3Effect}}
\end{figure}

For illustration, in Figure~\ref{J3Effect}, we draw the 
contour plots of the new Hamiltonian for different 
values of $J_3$ (or for different values of $\delta$). 
We see that when the $\delta$ coefficient increases 
(in absolute value), the vertical equilibrium goes down while 
the horizontal equilibrium goes below the 
``line $\sin\omega=0$''. We point out that, from some 
values of $\delta$, the equilibrium $\omega=-\pi/2$ 
disappears (Fig.~\ref{J3Effect} right panel).

\subsubsection{BepiColombo and other missions}

At present time, the semi-major axis of the two orbiters (MPO \& MMO) of 
the BepiColombo mission are respectively equal to $3\,394$ km and 
$8\,552$ km. The MPO altitude corresponds to our (E') phase space where 
the eccentricity vector has a circular concentric motion. 
The MMO initial conditions, without thrust correction, leads to a crash 
onto the Mercury surface after 3 years. 
Thanks to our theory, we can choose another initial condition
$a = 7\,355$ km and $e = 0.652$, that avoids the crash on 
Mercury and whose eccentricity vector is fixed.

\section{Conclusions}

The orbit dynamics of a space probe orbiting a planet or a natural 
planetary satellite has been investigated. The proposed model 
includes the effects of $J_2$ for the central body and the perturbation 
of the third body. We have developed a doubly averaged Hamiltonian 
and studied the location of the stable and unstable frozen orbits. 
Our analytical approach allows us to compute also the periods of 
the free librations at the equilibria. The analytical results 
have been checked and validated numerically by performing numerical 
integrations of the complete systems. Our theory is able 
to explain the behavior of our preliminary numerical investigations 
where the variation of the amplitude of the eccentricity is null 
and the presence of a separatrix has been found by numerical investigation. 
The theory is general enough to be applied to a wide range of probes 
around any planet or any natural planetary satellite, provided that 
they respect the hypotheses used to obtain our Hamiltonian model. 

We have shown the protection mechanism of the coefficient $J_2$ on the 
increasing of the eccentricity due to Kozai-Lidov effect. This 
mechanism is therefore able to find a larger number 
of frozen orbits than for the only Kozai-Lidov problem. 
We have also explained the asymmetry of the frozen 
equilibria caused by the addition of the coefficient $J_3$. 
We have also brought to the light a local deformation 
of the Kozai-Lidov equilibria that appears at high eccentricity, 
high inclination and large value of $\gamma$.

It would be interesting to take  this theory into account to choose the 
intial semi-major axis and eccentricity of an orbiter for future 
missions around planets or planetray satellites.

\begin{acknowledgements}
  The authors thank B.~Noyelles and A.~ Albouy 
  for fruitful discussions, the 
  \texttt{IMCCE} team for their welcome and B. Meyssignac for 
  initializing discussion.
  
  Numerical simulations were made on the local computing ressources 
  ({\it Cluster URBM-SYSDYN}) at the University of Namur (FUNDP, Belgium).
  
  This work was partly supported by the fellowship 
  {\it Concours des bourses de voyage de la Communauté Française de Belgique} 
  obtained by Nicolas Delsate.
\end{acknowledgements}




\begin{thebibliography}{18}
\providecommand{\natexlab}[1]{#1}
\providecommand{\url}[1]{{#1}}
\providecommand{\urlprefix}{URL }
\expandafter\ifx\csname urlstyle\endcsname\relax
  \providecommand{\doi}[1]{DOI~\discretionary{}{}{}#1}\else
  \providecommand{\doi}{DOI~\discretionary{}{}{}\begingroup
  \urlstyle{rm}\Url}\fi
\providecommand{\eprint}[2][]{\url{#2}}

\bibitem[{{Anderson} et~al(1987){Anderson}, {Colombo}, {Esposito}, {Lau}, and
  {Trager}}]{Anderson1987}
{Anderson} J, {Colombo} G, {Esposito} P, {Lau} E, {Trager} G (1987) The mass,
  gravity field, and ephemeris of mercury. Icarus Vol.71:pp.337--349

\bibitem[{{Brouwer} and {Clemence}(1961)}]{BrouwerClemence}
{Brouwer} D, {Clemence} G (1961) Methods of Celestial Mechanics. Academic Press

\bibitem[{{Garcia} et~al(2007){Garcia}, {de Pascale}, and {Jehn}}]{Garcia2007}
{Garcia} D, {de Pascale} P, {Jehn} R (2007) Bepicolombo mercury cornerstone
  consolidated report on mission analysis. Tech. rep., MAO Working Paper No.
  466, ESOC

\bibitem[{{Hairer} et~al(1993){Hairer}, {Norsett}, and {Wanner}}]{Hairer1993}
{Hairer} E, {Norsett} S, {Wanner} G (1993) Solving ordinary differential
  equations I. Nonstiff problems. 2nd edition. Springer-Verlag

\bibitem[{{Kozai}(1962)}]{Kozai1962}
{Kozai} Y (1962) Secular perturbations of asteroids with high inclination and
  eccentricity. Astronomical Journal Vol.67:pp.591

\bibitem[{{Laskar}(1988)}]{Laskar1988}
{Laskar} J (1988) {Secular evolution of the solar system over 10 million
  years}. Astronomy and Astrophysics 198:pp.341--362

\bibitem[{{Laskar}(2005)}]{Laskar2005}
{Laskar} J (2005) Hamiltonian systems and fourier analysis: new prospects for
  gravitational dynamics, Advances in Astronomy and Astrophysics, chap
  Frequency map analysis and quasiperiodic decomposition, pp 99--129

\bibitem[{{Lema{\^i}tre} et~al(2009){Lema{\^i}tre}, {Delsate}, and
  {Valk}}]{lemaitre2009}
{Lema{\^i}tre} A, {Delsate} N, {Valk} S (2009) {A web of secondary resonances
  for large A/m geostationary debris}. Celestial Mechanics and Dynamical
  Astronomy Vol.104:pp.383--402

\bibitem[{{Lidov}(1963)}]{Lidov1963}
{Lidov} ML (1963) Evolution of the orbits of artificial satellites of planets
  as affected by gravitational perturbation from external bodies. AIAA Journal
  p pp.1985

\bibitem[{{Paskowitz} and {Scheeres}(2004)}]{Paskowitz2004}
{Paskowitz} M, {Scheeres} D (2004) Orbit mechanics about planetary satellites.
  American Astronautical Society Vol.244

\bibitem[{{Paskowitz} and {Scheeres}(2006)}]{Paskowitz2006}
{Paskowitz} M, {Scheeres} D (2006) Design of science orbits about planetary
  satellites: Application to europa. Journal of Guidance, Control and Dynamics
  Vol.29

\bibitem[{{Saleh} and F.A.(2009)}]{Saleh2009}
{Saleh} L, FA R (2009) The stability and dynamics of planets in tight binary
  systems. The Astrophysical Journal Vol.694:pp.1566--1576

\bibitem[{{San-Juan} et~al(2006){San-Juan}, {Lara}, and {Ferrer}}]{SanJuan2006}
{San-Juan} J, {Lara} M, {Ferrer} S (2006) Phase space structure around oblate
  planetary satellites. Journal of Guidance, Control, and Dynamics Vol.29

\bibitem[{{Scheeres} et~al(2001){Scheeres}, {Guman}, and
  {Villac}}]{Scheeres2001}
{Scheeres} D, {Guman} M, {Villac} B (2001) Stabillity analysis of planetary
  satellite orbiters: Application to the europa orbiter. Journal of Guidance,
  Control and Dynamics Vol.24

\bibitem[{{Standish}(1998)}]{standish98}
{Standish} EM (1998) {JPL} planetary and lunar ephemeris, de405/le405. {JPL}
  Interoffice Memorandum IOM 312.D-98-048

\bibitem[{{Stoer} and {Bulirsch}(1980)}]{bulirsh-stoer}
{Stoer} J, {Bulirsch} R (1980) Introduction to numerical analysis.
  Springer-Verlag, New York

\bibitem[{{Sturm}(1835)}]{Sturm}
{Sturm} C (1835) Mémoire présentés par divers saavnts à l'Académie royale
  des SCiences de l'Institut de France, vol Vol.6, chap Mémoire sur la
  résolution des équations numériques

\bibitem[{{Tremaine} et~al(2009){Tremaine}, {Touma}, and
  {Namouni}}]{Tremaine2009}
{Tremaine} S, {Touma} J, {Namouni} F (2009) Satellite dynamics on the laplace
  surface. The astronomical journal Vol.1137:pp.3706--3717

\end{thebibliography}

\appendix
\section*{Appendix: {\it``Le théorème d'algèbre de Sturm''}}

Let $f(x)$ be a polynomial of positive degree with real coefficients and 
let $\{f_0(x),\, f_1(x),\, f_2(x),\,\dots,\, f_s(x)\}$ be the standard 
sequence for $f(x)$ such as 
\begin{eqnarray*}
  f_0(x) & = & f(x)\\
  f_1(x) & = & f'(x)\\
  f_2(x) & = & q_0(x)f_1(x) - f_0(x),\qquad deg\,f_2 < deg\,f_1\\
  f_3(x) & = & q_1(x)f_2(x) - f_1(x),\qquad deg\,f_3 < deg\,f_2\\
  \dots\\
  f_{i+1}(x) & = & q_{i-1}(x)f_i(x) - f_{i-1}(x),\qquad deg\,f_{i+1} < deg\,f_i\\ 
  \dots\\
  \text{until}\,f_{s+1}(x) & = & 0\,
\end{eqnarray*}
where $f_{i-1}$ is obtained by the Euclidean division: $f_{i+1}=q_{i-1}f_i-f_{i-1}$.
Assume $[a,b]$ is an interval such that $f(a)\neq 0 \neq f(b)$. Then the 
number of distinct roots of $f(x)$ in $[a,b]$ is $V_a-V_b$ where $V_c$ 
denotes the number of variations in sign of $\{f_0(c),f_1(c),\dots,f_s(c)\}$. 
The $0$ are dropped from the sequence.

\end{document}